\documentclass[aps,pra,twocolumn,groupedaddress,nofootinbib]{revtex4-1}

\usepackage{graphicx}
\usepackage{amsmath}
\usepackage{relsize}
\usepackage{color,soul}
\usepackage{textcomp}
\usepackage{tikz}
\usepackage{pgfplots}
\usepackage[bottom]{footmisc}
\usepackage{cleveref}
\usepackage{soul}
\pgfplotsset{compat=1.3}
\pgfplotsset{colormap={jet}{rgb255(0)=(0,0,143);rgb255(8)=(0,0,255);rgb255(24)=(0,255,255);rgb255(40)=(255,255,0);rgb255(56)=(255,0,0);rgb255(64)=(128,0,0)}}

\newcommand{\ket}[1]{|#1 \rangle}
\newcommand{\bra}[1]{\langle #1|}
\newcommand{\braket}[2]{\langle #1| #2\rangle}

\begin{document}

\title{Linear response theory of Josephson junction arrays in a microwave cavity}

\author{Samuel A. Wilkinson}
\affiliation{Chemical and Quantum Physics, School of Science, RMIT University, Melbourne, Victoria 3001, Australia}
\author{Jared H. Cole}
\affiliation{Chemical and Quantum Physics, School of Science, RMIT University, Melbourne, Victoria 3001, Australia}

\date{\today}

\begin{abstract}
Recent experiments on Josephson junction arrays (JJAs) in microwave cavities have opened up a new avenue for investigating the properties of these devices while minimising the amount of external noise coming from the measurement apparatus itself.
These experiments have already shown promise for probing many-body quantum effects in JJAs.
In this work, we develop a general theoretical description of such experiments by deriving a quantum phase model for planar JJAs containing quantized vortices.
The dynamical susceptibility of this model is calculated for some simple circuits, and signatures of the injection of additional vortices are identified.
The effects of decoherence are considered via a Lindblad master equation.
\end{abstract}

\maketitle

\section{Introduction}
The Josephson junction is one of the most important elements in turning quantum phenomena into usable technology. Both in existing devices such as SQUID magnetometers \cite{Fagaly2006}, the Josephson voltage standard \cite{Jeanneret2009,Hamilton1990} or superconducting filters \cite{Mitchell2016}, and in emerging technologies such as quantum computers \cite{Wendin2007,Makhlin2001}, quantum information processing devices \cite{Burkard2004,Zagoskin2011}, and Josephson metamaterials \cite{Anlage2011,Rakhmanov2008,Zhang2017}, the Josephson junction acts as a bridge between quantum mechanics at the micro-scale and practical technologies at the meso- and macro-scale.

The interest in Josephson junctions is not purely technological - they are also of interest from a  fundamental perspective. Large arrays of Josephson junctions act approximately as realisations of well-studied theoretical models such as the XY, Bose-Hubbard and sine-Gordon models \cite{Fazio2001}, which makes them excellent systems for studying quantum and classical phase transitions \cite{Cedergren2017a,Choi1998} and topological excitations such as vortices and single charge solitons \cite{Haviland1996,Homfeld2011,Ammanl1992,Oudenaarden1996}.
The fabrication technology for these systems is sufficiently advanced that the parameters governing the physics of interest can be selected with a very high degree of precision, and as such they can serve as model systems for investigating mesoscopic transport phenomena.

The behaviour of Josephson junction devices varies greatly as one moves through parameter space, which makes their study challenging.
However, we can exploit the so-called ``self duality" of the junction, which maps the weakly-interacting sector to the strongly-interacting sector and thereby makes the problem tractable \cite{Blanter1997,VanWees1991}. This will serve as an important tool in the analysis to follow.

Both fundamental investigations and technological applications of Josephson junction devices are limited by the ubiquity of charge noise and disorder.
It has been shown that the presence of charge disorder can qualitatively change the transport properties of Josephson junction arrays \cite{Vogt2015a,Vogt2015,Walker2015}, and mitigating charge noise is a key design criterion in the development of a superconducting quantum computer \cite{Astafiev2004, Schreier2008}.
If one wishes to probe a Josephson junction device experimentally, one typically attaches normal-conducting leads and performs transport experiments.
While much has been learned from this approach, it inevitably adds an additional source of charge noise and drives the system far from equilibrium.
Experiments that avoid external leads are therefore desirable.

\begin{figure}[tb!]
	\includegraphics[width=0.8\linewidth]{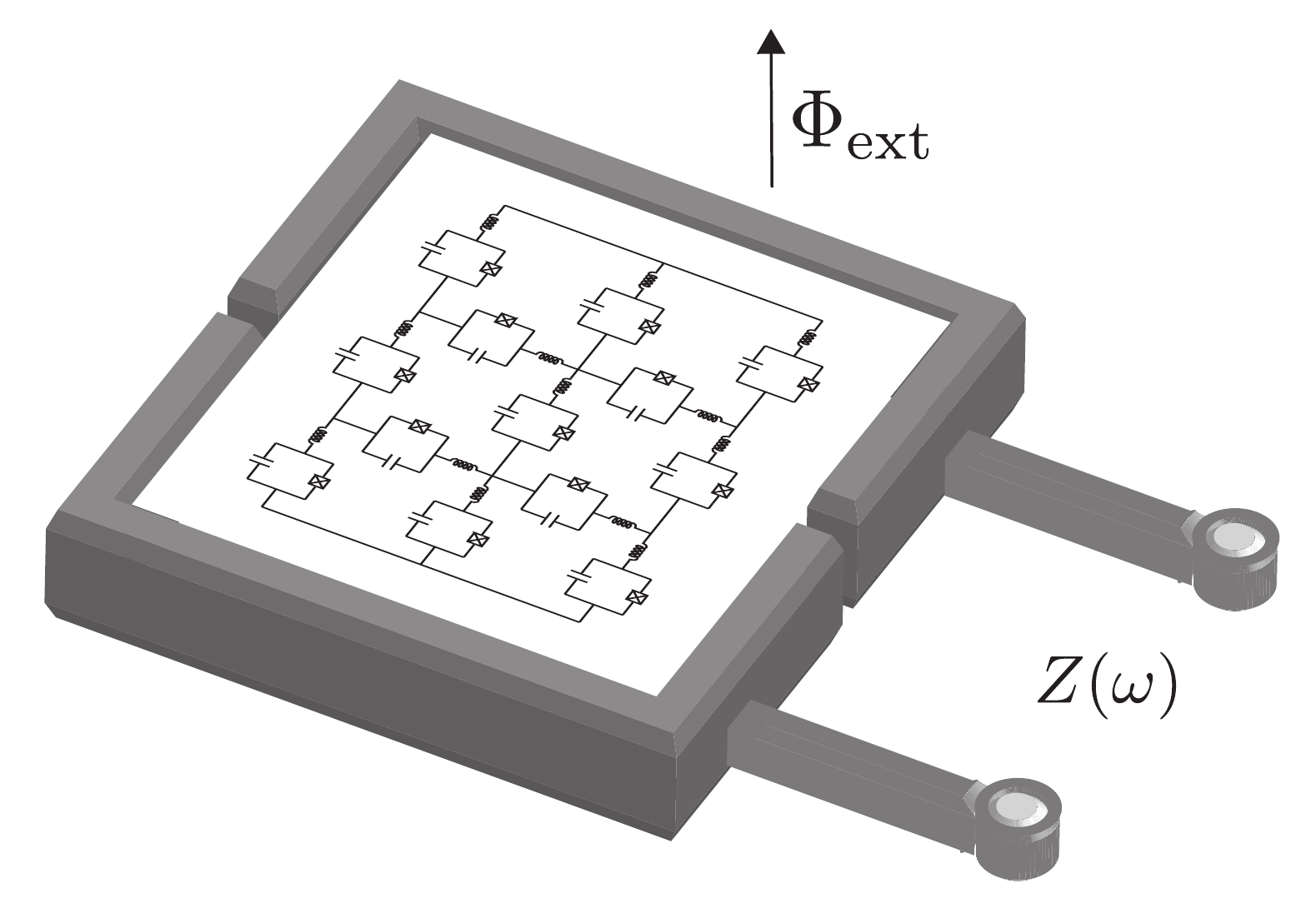}
	\caption{ \label{fig:Cavity} Cartoon of a Josephson junction array inside a microwave cavity. Such set-ups have been used experimentally to probe quantum behaviour of Josephson junction devices via spectroscopy measurements. }
\end{figure}

Motivated by the success of the 3D transmon qubit \cite{Paik2011}, several experimental groups have begun investigating Josephson junction devices by placing them inside microwave cavities \cite{Cosmic2018a,Chiena,Rastelli2018}, as is schematically illustrated in Fig.~\ref{fig:Cavity}.
This allows the device to be probed via spectroscopy, rather than transport measurements.
The effects of charge noise from external sources should be minimised in such experiments, and transport should be close to equilibrium.

Here we develop a theoretical framework for modelling these spectroscopic experiments on Josephson junction devices.
We first establish methods for obtaining Hamiltonian descriptions of the devices, and then use linear response theory to extract quantities that may be measured experimentally.
We pay particular attention to the admittance and impedance, and note that from these the single-port scattering parameter $S_{11}$ can be readily obtained.
	
The dynamics of Josephson junction arrays is largely governed by two energy scales: the charging energy $E_C$ required to add an additional charge to an island, and the Josephson energy $E_J$, associated with tunnelling of charges across a junction.  Utilizing duality arguments, we will develop equivalent approaches for calculating the impedance of charge-dominated arrays with $E_C \gg E_J$ and for calculating the admittance of flux-dominated arrays with $E_C \ll E_J$. 
The physics of the flux-dominated JJ arrays is dual to that of superconducting networks consisting of coherent quantum phase slip elements \cite{Ulrich2016,Mooij2006a,Kerman2013a,Kadin1990} instead of Josephson junctions, and as such the theory developed here will also be applicable to those devices.

\section{Model building}

The passage from circuit diagram to Hamiltonian is usually undertaken with a node- or edge-flux approach \cite{Devoret1997,Vool2017}. 
In the node-flux approach, we obtain a Lagrangian description in terms of the flux $\Phi$ associated with each node, and then perform a Legendre transformation to arrive at a Hamiltonian description in terms of the phase of the superconducting condensate wavefunction $\phi = 2\pi\Phi/\Phi_0$ and the number of Cooper-pairs $n$ on each node of the circuit (or, alternatively, differences in phase and Cooper-pair number across the branches of the circuit).
This can readily be quantized by imposing the canonical commutation relation $[\phi_i, n_j] = i\delta_{ij}$, where $i$ and $j$ label different nodes (or, alternatively, different branches) of the circuit.
This process will lead us to a description in terms of discrete Cooper-pairs on a lattice, which may tunnel from site to site via Josephson junctions.

If our circuit is planar, however, there is an alternative approach we may take based on loop-charges \cite{Ulrich2016}.
This approach is a Lagrangian/Hamiltonian formalisation of the mesh analysis which is commonplace in electrical engineering \cite{Pointon1991}.
We begin by obtaining a Lagrangian defined in terms of so-called ``loop charges" $Q$, each one defined within a plaquette (irreducible loop) of the circuit.
These loop charges are fictitious in that they do not correspond to any physical observable, but the difference $Q_i - Q_j$ between two adjacent loop charges corresponds to the charge polarization across the branch common to the two loops.
As with the node/branch flux, we can get a quantum description first by taking the Legendre transformation and then by imposing a canonical commutation relation.
In this case, we demand that $[Q_i,\Phi_j] = i\hbar\delta_{ij}$ where $\Phi_j$ is the flux threading the $j^\textrm{th}$ loop.

The main shortcoming of the loop-charge approach is the restriction to planar circuits.
This makes it difficult to include a ground plane in the description.
However, in 2D devices in which the ground capacitance is negligible, the loop-charge approach offers a convenient way to study arrays in the limit where vortices are the relevant single-particle excitations.

The node/branch-flux approach is appropriate for small-capacitance junctions, where charging effects dominate, i.e. when $E_C > E_J$. 
For a generic 2D array consisting only of nodes connected to each other via Josephson junctions with Josephson energy $E_J$ and capacitance $C_J$, and to a ground plane via a capacitance $C_G$, the node-flux approach ultimately produces a quantum phase model Hamiltonian \cite{Fazio2001}
\begin{equation} \label{eq:quantumPhaseModel}
	H = \frac{(2e)^2}{2}\sum_{ij}\left(n_i - \tilde{n}_i\right)C_{ij}^{-1}(n_j - \tilde{n}_j) - E_J \sum_{\langle ij \rangle}\cos(\phi_i - \phi_j).
\end{equation}
where $\tilde{n}_i$ is the effective charge on the $i^\textrm{th}$ island due to external gate voltages or charge disorder.

In the opposite limit, ground capacitance $C_G$ of the superconducting islands is large and single-charge effects can be neglected due to the smallness of $E_C = (2e)^2/2C_g$. Here, charge ceases to be a good quantum number and instead the effects of single flux quanta become important. To study this limit, we employ the loop-charge approach to derive an equivalent dual circuit.

\subsection{Vortex lattice model}
A limitation of the loop-based approach is the inability to handle non-linear inductors, which would make it seem a poor choice for the modelling of an array of Josephson junctions (which are close to as non-linear an inductor as one can find). However, we will show that by beginning in a mixed representation and integrating out fast-moving variables, we can transform the model from one of a lattice of non-linear inductors (Josephson junctions) to one of non-linear capacitors [e.g coherent quantum phase slip (QPS) elements]. It is important to note that this transformation is purely one of description - the physical system remains a Josephson junction array.

\begin{figure}[tb!]
	\includegraphics[width=\linewidth]{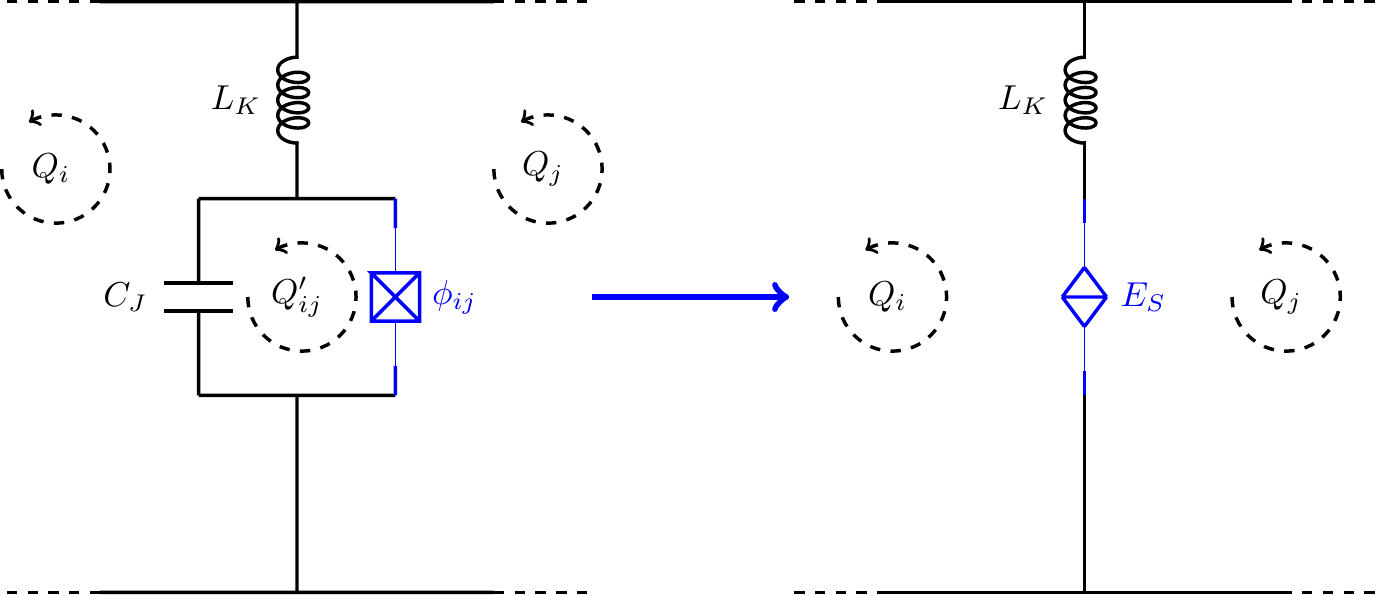}
	\caption{ \label{fig:Shunted JJ} Illustration of the mixed-representation, which allows us to incorporate Josephson junctions into a loop-charge approach, and the Born-Oppenheimer approximation which allows us to replace the capacitively-shunted JJ with an effective QPS element. The branch with the Josephson junction (coloured blue) is initially considered within the node-flux representation, while all other branches are incorporated via the loop-charge representation. The fictitious loop charge $Q'_{ij}$ will be reduced to an algebraic constraint. When fast-moving degrees of freedom are integrated out, the Josephson junction is approximated by an effective quantum phase slip element. 
	}
\end{figure}

The mixed approach is depicted in Fig~\ref{fig:Shunted JJ}, where the branch with the Josephson junction (in blue) is treated using the branch variable $\phi_{ij}$ (thus the non-linear part of the Josephson junction is not treated in the loop-charge formalism). Fig.~\ref{fig:Shunted JJ} shows a single junction represented in the capacitively shunted junction model. A realistic junction also includes a kinetic inductance $L_K$ \cite{Agren2001}, which we have represented here as running in series with the the tunnel junction, as well as a geometric inductance $L_G$ associated with each loop.

An additional limitation of the loop-charge approach is the restriction to \textit{planar} circuits \cite{Ulrich2016}.
Thus, in the following derivation, we will not include a ground plane (which would violate the planarity of the circuit and by extension the validity of the loop charge approach).
This is equivalent to assuming that charging effects will be small.
In a cavity set-up this should indeed be true as the array should be be quite far from the walls of the cavity the capacitance between the circuit and the ground should be negligible.
If, however, we wished to include a ground plane, this task would be more complicated.
Obtaining a circuit theory Lagrangian may still be possible via a mixed representation, however it will introduce additional degrees of freedom.

We will now derive a description of a 2D JJ array in terms of loop-based degrees of freedom.
Each irreducible loop of the circuit is assigned a loop charge $Q$.
When two loops share a common branch, then a term appears in the Lagrangian depending on the circuit element contained in that branch.
A capacitance $C$ between loops 1 and 2 will contribute a term $(Q_1 - Q_2)^2/2C$, while an inductance $L$ will contribute $L(\dot{Q}_1 - \dot{Q}_2)^2/2$.
Furthermore, each loop with have an associated geometric inductance $L_G$, which contributes $L_G\dot{Q}^2/2$.
If there is an external magnetic flux $\Phi_\textrm{ext}$, this will provide the Lagrangian with an addition term $\Phi_\textrm{ext}\dot{Q}$ for every irreducible loop in the circuit.

There are, however, further complications arising from the mixed representation.
In Fig.~\ref{fig:Shunted JJ}, we see that the JJs are to be treated in a node-flux approach, and thus they mark the boundary between node-flux and loop-charge representations.
Each such boundary contributes a term $\left(Q_1 - Q_2 \right)\left(\dot{\phi}_i - \dot{\phi}_j\right)$, where $Q_1$ and $Q_2$ are the loop charges on either side of this boundary branch and $\phi_1$ and $\phi_2$ are node fluxes at either end of the boundary branch.

Assuming a square lattice geometry where each loop has geometric inductance $L_G$, each branch has inductance $L_K$ (here assumed to be a kinetic inductance, it may also have contributions from the Bloch inductance of the JJs), and a JJ with Josephson energy $E_J$ and capacitance $C_J$ (we assume self-capacitance to be negligible).
The branch flux through a JJ is given by $\phi_{ij} = \phi_j - \phi_i$.
This gives us a Lagrangian
\begin{widetext}
\begin{equation}
\begin{split}
\mathcal{L}_\textrm{lattice}[Q,\dot{Q},\phi,\dot{\phi},Q'] =& \sum_{\langle i,j \rangle} \left[\frac{ \left(Q'_{ij} - Q_i \right)^2}{2C_J} + \frac{L_K}{2}\left(\dot{Q}_i - \dot{Q}_j\right)^2 + E_J\cos\left(\frac{2\pi}{\Phi_0}\phi_{ij} \right) - \left(Q_{j} - Q'_{ij}\right)\dot{\phi}_{ij} \right]\\
&+ \sum_i \left[\frac{L_G}{2}\dot{Q}_i^2 + \Phi_\textrm{ext}\dot{Q}_i \right]
\end{split}
\end{equation}
\end{widetext}
Since derivatives of $Q'$ do not appear in the Lagrangian (the fictitious loop has no inductance), the Euler-Lagrange relations for this variable yield only the algebraic constraint
\begin{equation}
Q'_{ij} = Q_i - C_J\dot{\phi}_{ij}.
\end{equation}
We can therefore write the Lagrangian
\begin{widetext}
\begin{equation}
\begin{split}
\mathcal{L}_\textrm{lattice}[Q,\dot{Q},\phi,\dot{\phi}] =& \sum_{\langle ij \rangle} \left[ \frac{1}{2}\dot{Q}_i L_{ij} \dot{Q}_j + \left(Q_i - Q_j\right)\dot{\phi}_{ij} + \frac{1}{2}C_J\dot{\phi}_{ij}^2 + E_J\cos\left(\frac{2\pi}{\Phi_0}\phi_{ij}\right) \right] + \Phi_\textrm{ext} \sum_i \dot{Q}_i
\end{split}
\end{equation}
\end{widetext}
where $L_{ij}$ is the inductance matrix,
\begin{equation}
L_{ij} = \begin{cases} z_iL_K + L_G \quad & i = j\\
-L_K \quad & j \in \mathcal{N}(i)\\
0 \quad & \textrm{otherwise} 
\end{cases}
\end{equation}
Here $z_i$ is the co-ordination number of site $i$ ($z_i$ = 4 for all sites on a square lattice) and $\mathcal{N}(i)$ is the neighbourhood of site $i$.

It will be convenient at this point to introduce the vector notation $\vec{Q} = (Q_1, Q_2, \dots Q_N)^\textrm{T}$.

We move to a Hamiltonian description by obtaining the conjugate variables
\begin{equation}
q_{ij} = \frac{\partial \mathcal{L}}{\partial \dot{\phi}_{ij}} = Q_i - Q_j + C_J\dot{\phi}_{ij}; \quad \vec{\Phi} = \frac{\partial \mathcal{L}}{\partial \dot{\vec{Q}}} = \textbf{L}\dot{\vec{Q}} + \Phi_{ext}\vec{1}
\end{equation}
where $\vec{1}$ is the vector of length $N$ whose elements are all 1. For notational convenience, we will write $\Phi_{ext}\vec{1} = \vec{\Phi}_{ext}$. $\vec{\Phi} = (\Phi_1,\Phi_2,\dots,\Phi_N)^\textrm{T}$ is a vector of loop flux operators which are conjugate to the loop charge operators.

Our Hamiltonian is then
\begin{equation}
\begin{split}
\mathcal{H} =& \frac{1}{2} \left(\vec{\Phi} - \vec{\Phi}_\textrm{ext}\right)^\textrm{T}\textbf{L}^{-1}\left(\vec{\Phi} - \vec{\Phi}_\textrm{ext}\right)\\ &+ \sum_{\langle i,j\rangle}\left[ \frac{\left(q_{ij} - Q_i + Q_j\right)^2}{2C_J} - E_J\cos\left(\frac{2\pi}{\Phi_0}\phi_{ij}\right)\right].
\end{split}
\end{equation}
The terms inside the sum can be readily recognised as a sum of single junction Hamiltonians, where $Q_i - Q_j$ plays the role of the quasicharge \cite{Likharev1985}. If we assume that the junction variables $q_{ij}$ and $\phi_{ij}$ evolve quickly compared with the loop variables $Q_i - Q_j$, we can employ a Born-Oppenheimer approximation and diagonalize the single junction Hamiltonians with respect to a fixed, classical value of $Q_i - Q_j$.

So long as the circuit is driven adiabatically, we can take the lowest energy band of the single junction Hamiltonian as an effective periodic potential felt by the loop degrees of freedom. This approximation is identical to the quasicharge approach that has been used in the study of single junctions \cite{Likharev1985} and linear arrays \cite{Vogt2015a,Haviland1996, Wilkinson2017}.

The energy bands of a single Josephson junction are given by the characteristic values of Mathieu's equation \cite{Wilkinson2018}, where the quasicharge $Q_i - Q_j$ plays the role of the Floquet exponent. In the limit that $E_J/E_C \gg 1$, the lowest energy is approximately a cosine of the quasicharge. Inserting this into the Hamiltonian, we find
\begin{equation} \label{eq:LoopHam}
\begin{split}
\mathcal{H} =& \frac{1}{2} \left(\vec{\Phi} - \vec{\Phi}_\textrm{ext}\right)^\textrm{T}\textbf{L}^{-1}\left(\vec{\Phi} - \vec{\Phi}_\textrm{ext}\right)\\ &- E_S \sum_{\langle i,j \rangle} \cos\left(\frac{Q_i - Q_j}{2e}\right)
\end{split}
\end{equation}
where
\begin{equation}\label{eq:ESdef}
E_S = 32 \left(\frac{E_JE_C}{\pi}\right)^{1/2} \left(\frac{E_J}{2E_C}\right)^{1/4} \exp\left[-\left(8\frac{E_J}{E_C}\right)^{1/2}\right].
\end{equation}
At a glance, it may seem as if capacitance has disappeared from the problem.
However, the self-capacitance was neglected initially as as we consider a system where the walls of the cavity are far from the array itself. The junction capacitance $C_J$ has been absorbed into $E_S$ as given in Eq.~ \ref{eq:ESdef}.

The Born-Oppenheimer approximation assumes that the system is always in the ground state with respect to the fast-moving degrees of freedom (Cooper pairs).
Under this assumption, the quasicharge becomes a periodic variable with period $2e$, because changing the quasicharge by $\pm e$ will simply cause a Cooper pair to tunnel across a junction so as to remain in the ground state.
Because $Q$ is now compact, its canonical conjugate $\Phi$ becomes discrete.
This can be understood heuristically by noting that we have effectively replaced a Josephson junction with a coherent quantum phase slip (QPS) element (as evidenced by the $\cos(Q)$ term in the Hamiltonian).
If this replacement is taken literally, we now have an uninterrupted superconducting loop, so that the flux through it becomes quantized.
In two-dimensional arrays with large $E_J$, these flux quanta manifest as vortices, and we will therefore refer to them as vortices here.
To make this approximation explicit, we will draw JJs as QPS elements in circuit diagrams which we treat under the Born-Oppenheimer approximation.

Expressed in the vortex-number basis, the second term in the Hamiltonian becomes
\begin{equation} \label{eq:hoppingTerm}
-\frac{1}{2}E_S \sum_{\langle i,j \rangle} \sum_{n,m} \left( \ket{n_i+1,m_j-1}\bra{n_i,m_j} + \textrm{H.c.} \right)
\end{equation}
where $n_i (m_j)$ label the number of vortices on site $i$ $(j)$.

The replacement of a Josephson junction with a QPS element is a consequence of the so-called self duality of a Josephson junction, and an example of the electromagnetic duality between charge and flux in electrical circuits.
For any circuit consisting of Josephson junctions, there is a dual circuit consisting of QPS elements \cite{Mooij2006a}.
In particular, the model in Eq.~\ref{eq:LoopHam} is an exact dual to the usual quantum phase model for a Josephson junction array expressed in terms of island charges and fluxes, given in Eq.~\ref{eq:quantumPhaseModel}.
Thus for every vortex-based circuit described by Eq.~\ref{eq:LoopHam}, there is a dual charge-based circuit described by Eq.~\ref{eq:quantumPhaseModel}.

We will present most of this work in the vortex language, but there is a simple translation between vortex-based circuits and charge-based circuits.
Circuit diagrams for flux-based circuits will be drawn with QPS elements in place of JJs, with the understand that they equivalently represent Josephson junctions in the quasicharge limit discussed above.

\subsection{Classical limit}

In the limit $E_S \rightarrow 0$, tunnelling is suppressed and the system becomes a classical lattice of fluxes. 
Finding the ground state is simply a matter of energy optimisation.
At zero external flux this is trivial: the ground state is the state with no fluxes at all in it.

As the external flux is increased, we will inject more fluxes into the array.
A simple calculation shows that the state containing a single vortex at site $k$ is lower in energy than the empty state when the external frustration reaches 
\begin{equation}
f = \frac{L_{kk}^{-1}}{2\sum_i L_{ik}^{-1}}.
\end{equation}

For a completely homogeneous system, the exact value of the index $k$ is completely arbitrary.
When a boundary is included, however, the situation is different as $L_{kk}$ will vary across the array.
$L_{kk}$ will be lowest towards the centre of the array, so that is where the first vortex will appear.

In considering the appearance of two vortices at higher frustrations, we need to be careful where they appear.
They will want to avoid edges of the array much like the single vortex did, but they will also want to avoid each other.
So we find a transition from the state of a single vortex at site $k$ to a state of two vortices at sites $q$ and $q'$ will occur at a frustration of
\begin{equation}
f = \frac{L^{-1}_{qq} + L^{-1}_{q'q'} + 2L_{qq'} - L_{kk}}{2\sum_i(L^{-1}_{iq} + L^{-1}_{iq'} - L^{-1}_{ik})}.
\end{equation}
Similar arguments apply as we increase the external flux, but as we do so the particular dimensions of the array become more and more important, and it is much more convenient to just calculate this numerically.
We eventually arrive at a completely full array at a frustration of 
\begin{equation}
f = 1 - \frac{L_{kk}^{-1}}{2\sum_i L_{ik}^{-1}}
\end{equation}
and therefore the width of the flux injection region is
\begin{equation} \label{eq:fluxInjectionWidth}
\Delta f = 1 - \frac{L_{kk}^{-1}}{\sum_i L_{ik}^{-1}}.
\end{equation}
In the experimental data of \cite{Cosmic2018a}, we see that $\Delta f$ approaches 1, meaning that the on-site inductive energy is not much larger than the inductive interaction between different sites.
In contrast, in the limit of negligible inductive interactions (so that the on-site interaction is dominant), $\Delta f$ approaches 0, so instead of a gradual injection of one vortex after another we get a steep, sharp injection of $N$ vortices at once (where $N$ is the number of plaquettes).

\section{Linear response theory}

When we probe a JJA in a microwave cavity with radiation, we are able to measure the electromagnetic response of the system via spectroscopy, rather than transport measurements.
To theoretically model this we assume coupling to the microwave radiation is relatively weak, so that it can be treated in linear response.

The response of the system to a time-dependent perturbation is given by the susceptibility, which may be calculated via the Kubo formula \cite{Kubo1970}
\begin{equation} \label{eq:Kubo}
\begin{split} 
	\chi_{\Phi}(t-t') =& -i\langle[\Phi(t),\Phi(t')]\rangle \theta(t-t') \\
	\chi_{\Phi}(\omega) =& \mathcal{F}[\chi_{\Phi}(t)]
\end{split}
\end{equation}
where $\theta(t)$ is the Heaviside step function, which enforces causality, and $\mathcal{F}$ is a Fourier transform.
At sufficiently low temperatures, the average $\langle \dots \rangle$ will simply be the ground state expectation value.
The time evolution of the $\Phi$ operators is calculated in the Heisenberg picture $\Phi(t) = e^{-iHt}\Phi e^{iHt}$.


The charge susceptibility $\chi_{Q}$ is given by a formula exactly analogous to $\chi_\Phi$.
The ultimate response functions of interest are the electrical impedance $Z$, and the admittance $Y$.
The impedance is defined by
\begin{equation}
\begin{split}
	\langle V(t) \rangle =& \langle V \rangle_0  \int_{-\infty}^t Z(t-t')I(t)dt'\\
	Z(\omega) =& \frac{V(\omega)}{I(\omega)}
\end{split}
\end{equation}
and the admittance is analogously defined through
\begin{equation}
\begin{split}
I(t) =& \int_{-\infty}^t Y(t-t')V(t)dt'\\
Y(\omega) =& \frac{I(\omega)}{V(\omega)}
\end{split}
\end{equation}
so that, trivially, $Z = Y^{-1}$.

Using the electromotive force formula $\langle V(t) \rangle = -d\langle \Phi \rangle/dt$, we see that
\begin{equation}
\begin{split}
\langle V(t) \rangle =& \int_{-\infty}^t \frac{d\chi_{\Phi}(t-t')}{dt}I(t')dt'\\
Z(t) =& \frac{d\chi_\Phi}{dt}\\
Z(\omega) =& i\omega\chi_\Phi(\omega).
\end{split}
\end{equation}
Similar reasoning, using the definition of current as $I = dQ/dt$ gives us
\begin{equation}
Y(\omega) = i\omega\chi_Q(\omega).
\end{equation}
These each give us the impedance/admittance of a single site in our system.


In the absence of dissipation, the zero-temperature response function is given by
\begin{equation} \label{eq:specRep}
	\chi_{A}(\omega) = \sum_{n} |\bra{\psi_n}A\ket{\psi_0}|^2 2\pi\delta(\omega - \omega_{n0})
\end{equation}
where $\omega_{n0}$ is the gap between the energy $E_n$ of the state $\ket{\psi_n}$ and the ground state energy $E_0$.

When considering an open system which may be in a mixed state, this formula must be modified slightly, as the correlator $\langle A(t)A(t')\rangle$ is now a weighted average over several states rather than a ground state expectation value.
The steady state of the system can be described by a density matrix $\rho = \sum_j w_j\ket{\psi_j}\bra{\psi_j}$, where $w_j$ are the statistical weights of the mixture.
In this case the non-dissipative response becomes
\begin{equation} \label{eq:mixedSpecRep}
\chi_{A}(\omega) = \sum_{n,m} w_m|\bra{\psi_n}A\ket{\psi_m}|^2 2\pi\delta(\omega - \omega_{nm})
\end{equation}
where $m$ runs over the states appearing in the steady state $\rho$.

\section{Two-site system}
We initially consider a system consisting of only two
loops connected by a tunnel junction. There are two different systems we can discuss here: the hard-boundary system depicted in Fig.~\ref{fig:2SiteClosed}, and the junction-boundary system depicted in Fig.~\ref{fig:2SiteOpen}, which
has tunnel junctions on the exterior so that particles
can enter and exit. Each these circuits has a dual which
obeys the same dynamical equations, as is discussed in
Appendix D . For clarity, we will initially restrict our attention to flux-based circuits, but the notion and much of
the discussion will be kept general so as to apply equally
well to their charge-based duals.

 each of which has a JJ form and an equivalent dual form in terms of QPS elements.
The charge-based circuit, Fig.~\ref{fig:2SiteOpen} a), is known as the double-island Cooper-pair box or superconducting SET \cite{Bibow2002,Lambert2014,Toppari2004}.

We can describe these circuits in a charge/vortex agnostic language by defining $\hat{n}_j$ as the number of particles, be they vortices or charges, on site $j$. $\hat{b}$ is the operator that reduces the number of particles by one, and $\hat{b}^\dagger$ increases the number of particles by one. (Note: these are not identical to the usual bosonic creation/annihilation operators, since $\hat{n}$ may have negative eigenvalues and thus cannot be written as $\hat{n} = \hat{b}^\dagger \hat{b}$. This technical point can be circumvented, but here we shall simply ignore it as it will not affect the physics of this simple system.)

\subsection{Hard boundary}

\begin{figure}

	\centering{a)} \includegraphics[width=5cm]{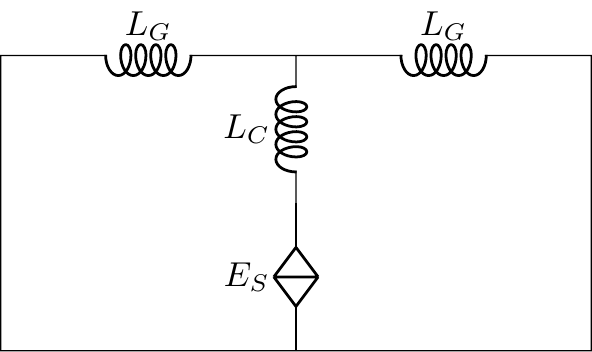}
	
	\vspace{0.8cm}
	
	\centering{b)}\includegraphics[width=\columnwidth]{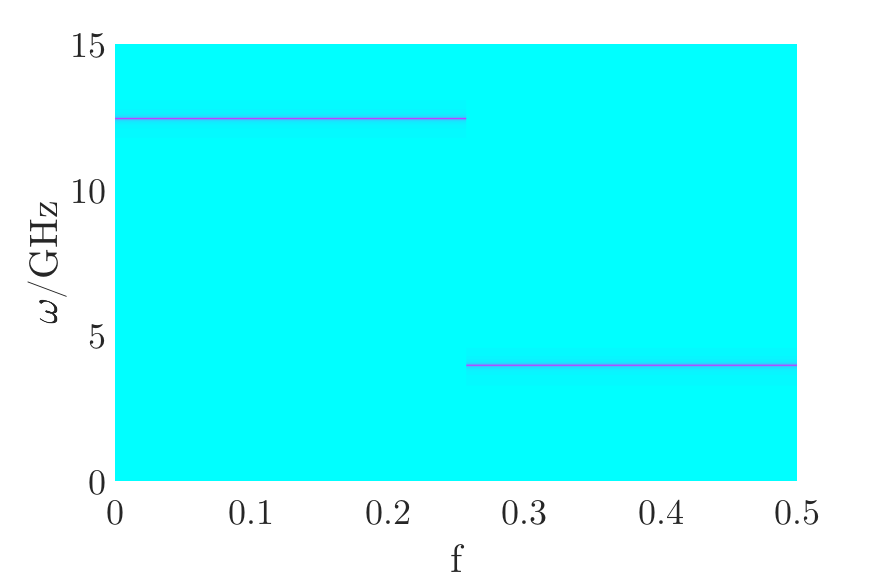}
	\caption{ \label{fig:2SiteClosed} a) A two-loop circuit with hard boundaries, so that vortices cannot enter or exit the circuit. b) The dynamical response $|\chi_{n}(\omega)|$ of the circuit depicted in a). The response consists only of sharp peaks at the resonance frequencies given in Eq.~\ref{eq:closedPeaks}. The susceptibility has been normalized at each value of $f$ to make peaks equally visible across the whole spectrum, so that the colour axis is arbitrary and does not represent the actual peak height. }
\end{figure}

We can write a charge/vortex-agnostic Hamiltonian
\begin{equation} \label{eq:twoSite}
\begin{split}
	H =& \frac{1}{2}\begin{pmatrix}
	\hat{n}_1 - f, & \hat{n}_2 - f
	\end{pmatrix}
	\begin{pmatrix}
	\beta & 1\\ 1 & \beta
	\end{pmatrix}
	\begin{pmatrix}
	\hat{n}_1 - f\\ \hat{n}_2 - f
	\end{pmatrix}\\
	& - \frac{t}{2}\left(\hat{b}_1^\dagger \hat{b}_2 + \hat{b}_1\hat{b}_2^\dagger\right)
\end{split}
\end{equation}
where we have written all energies in units where the off-diagonal inductive interaction strength is 1.
$t$ corresponds to the tunnelling amplitude (either $E_S$ or $E_J$), $\beta$ is the energy cost of adding a single particle to a site (i.e.\ the diagonal terms of the inductance or capacitance matrix), and $f$ is a generalized frustration. Note that, since the diagonal elements of the inverse inductance/capacitance matrix are always greater than the off-diagonal elements, $\beta \geq 1$.

If $t \ll 1$, we can restrict ourself to particle numbers of $n = 0$ and $n = \pm 1$.
With this restriction, the Hamiltonian is reduced to a $9\times 9$ matrix which may be diagonalized exactly.
The eigenstates and eigenvalues are given explicitly in Appendix~\ref{appendix:eigen}.
Using the labelling system given in that appendix, the ground at zero frustration is is $\ket{\psi_4}$.

The ground state changes character at a frustration of
\begin{equation}
	|f_c| = \frac{1 - t + \sqrt{(\beta - 1)^2 + 2t^2}}{2(\beta + 1)}.
\end{equation}
For $f > f_c$, the ground state is $\ket{\psi_8}$, and for $f < - f_c$, the ground state is $\ket{\psi_6}$

In the absence of dissipation, the zero-temperature linear response of this circuit is very simple.
Eq.~\ref{eq:specRep} can be calculated by noting that the matrix element will only be non-zero for states with the same number of excitations as the ground state.
In each of the three regimes ($|f| < f_c$, $f< -f_c$ and $f>f_c$) there is only one non-zero term.
We find that the reactive response of the system consists of sharp peak at the resonance frequency $\chi(\omega) \propto \delta(\omega - \omega_r)$,
\begin{equation} \label{eq:closedPeaks}
\begin{split}
\omega_r = \begin{cases}
 \omega_{7,6} = t,  & f < -f_c\\
 \omega_{1,4} = \frac{1}{2}\left[ \beta - 1 + \sqrt{(\beta - 1)^2 + 2t^2} \right], & |f| < f_c\\
 \omega_{9,8} = t, & f > f_c.
\end{cases}
\end{split}
\end{equation}
Note that within each region the response it completely independent of $f$.

The resulting response spectrum, calculated using Eq.~\ref{eq:Kubo}, can be seen in Fig.~\ref{fig:2SiteClosed}.
Except where stated otherwise, all calculations are performed with $E_S = 1$ GHz, $L_G = 10^{-3}$ nH and $L_K = 10^{-2}$ nH (or, equivalently, for charge-based circuits, $E_J = 1$ GHz, $C_G = 10^{-3}$ nF and $C_J = 10^{-2}$ nF).
The height of the peaks in $\chi(\omega)$ differ significantly, so a normalization has been applied to make the features easier to see \footnote{The normalization applied consists of dividing each value of $|\chi(\omega,f)|$ by the maximum value of $\chi(\omega)$ for that particular value of $f$, so that the maximum height of the peaks is always unity as $f$ tuned. Without these features, peaks at some values of $f$ are much larger than others.}.
For this reason, the colour axis is arbitrary, and this spectrum only gives information about locations of peaks and their \textit{relative} amplitudes at a given value of $f$.
The same normalization is applied to all other response spectra presented in this work.

\subsection{Junction boundary}

\begin{figure}
	\centering{a)} \includegraphics[width=5cm]{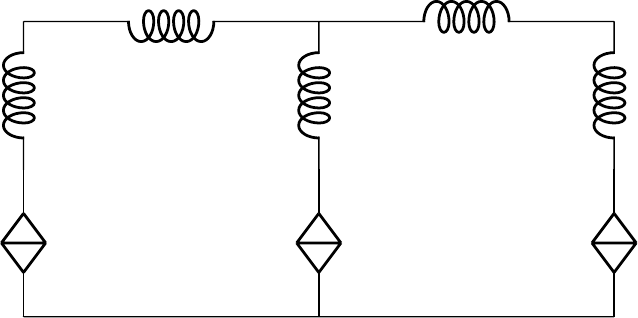}
	
	\vspace{0.8cm}
	
	\centering{b)}\includegraphics[width=\columnwidth]{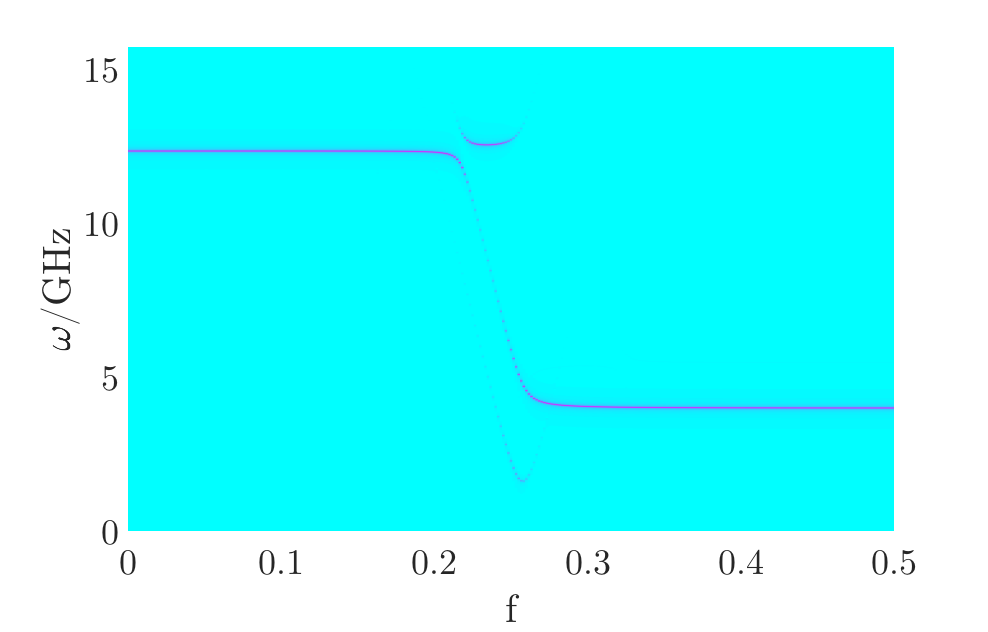}
	\vspace{0.8cm}
	\caption{\label{fig:2SiteOpen} a) A two-loop circuit with with junction boundaries, such that particle number is no longer conserved. b) The dynamical response $|\chi_{n}(\omega)|$ of the two-loop circuit depicted in a). Since particle number is no longer conserved, the zero-particle and one-particle ground states are adiabatically connected. Energy levels curve as they approach the crossover, and additional resonant frequencies appear when compared with the response in Fig.~\ref{fig:2SiteClosed}  }
\end{figure}

The problem becomes more interesting if we place additional tunnel junctions in the system, as depicted Fig.~\ref{fig:2SiteOpen}.
These add a term to our Hamiltonian
\begin{equation} \label{eq:boundary}
\hat{V} =	- \frac{t_\textrm{edge}}{2}\left( \hat{b}_1 + \hat{b}_1^\dagger + \hat{b}_2 + \hat{b}_2^\dagger  \right)
\end{equation}
which breaks conservation of particle number.
States can now exist in superpositions of different numbers of particles, and rather than having the ground state expectation value $\langle N \rangle = \sum_i \langle n_i \rangle $ change in sharp jumps at a particular value of $f$, we have a more gradual crossover to states of different total particle number.

With the edges open, an exact analytic solution is no longer accessible.
However, we can still numerically calculate the response of the system, obtaining the spectrum presented in Fig.~\ref{fig:2SiteOpen}.
In that calculation, we take $t_\textrm{edge} = t$.

The smooth crossover region can be clearly seen from Fig.~\ref{fig:2SiteOpen}. 
In this region, the average number of particles in the ground state is not an integer, as the ground state is not an eigenstate of the total particle number operator.
The width of this region can be estimated from the eigenspectum of the solvable circuit with closed edges.
We assume that the crossover in the open circuit begins when the ground state and first excited state of the closed system have a different number of particles, as this is when states of different particle number in the open circuit will begin to hybridize.

\begin{figure}
	\includegraphics[width=8cm]{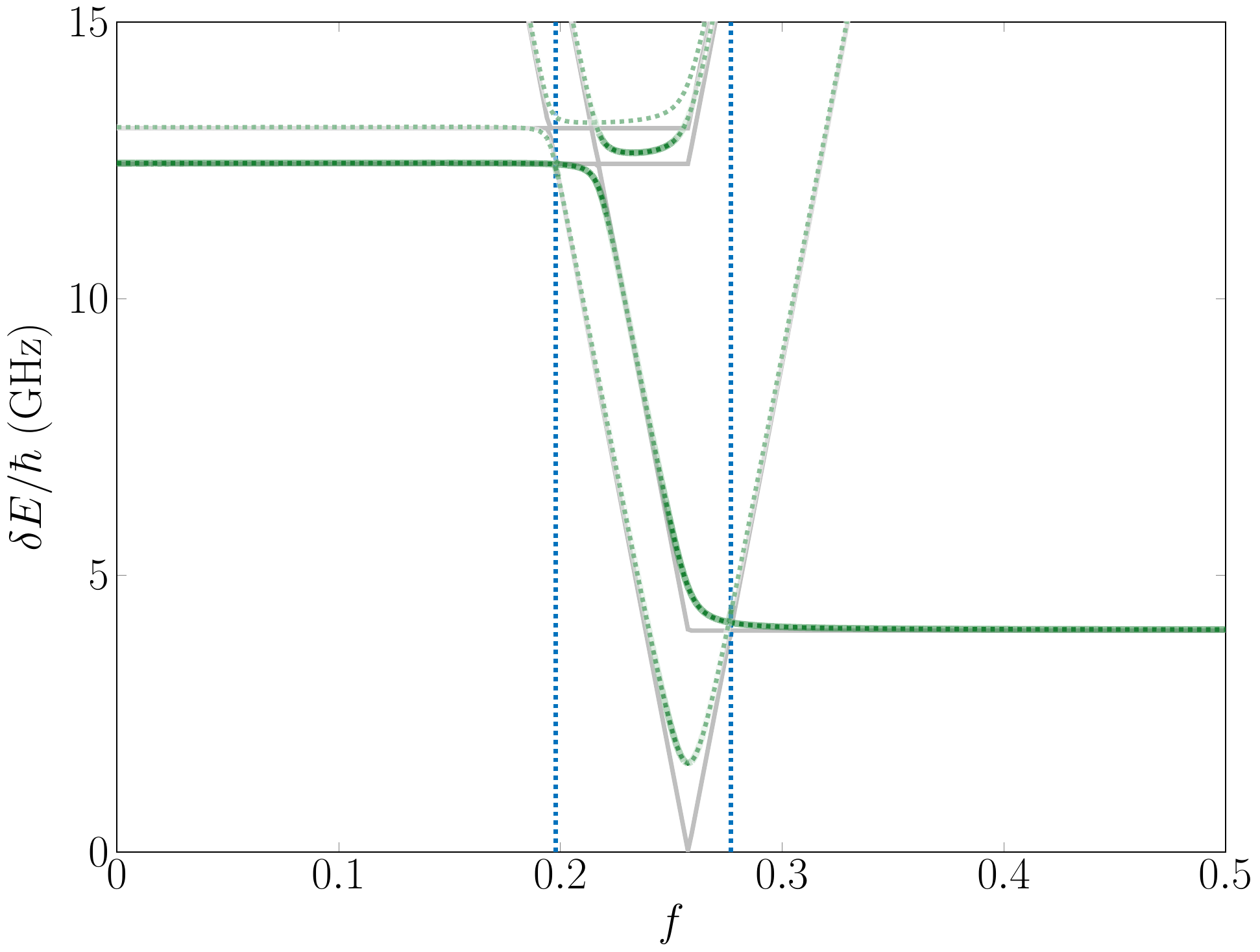}
	\centering
	\caption{ \label{fig:2x1overlay}  Energy gaps for a 2-site circuit with closed boundary (grey) and open boundary (dashed green). Vertical blue lines illustrate estimates for width of crossover region given by Eq.~\ref{eq:crossBegin} and \ref{eq:crossEnd}. The colour of the thick green line indicates the magnitude of the matrix element $|\bra{\psi_m}\hat{n}\ket{\psi_n}|$, which gives the magnitude of the linear response in accordance with Eq.~\ref{eq:specRep}. }
\end{figure}

This crossover begins when $\lambda_1 = \lambda_6$, at
\begin{equation} \label{eq:crossBegin}
f = \frac{1 - \frac{1}{2}(\beta + t)}{1 + \beta}
\end{equation}
and ends when $\lambda_4 = \lambda_7$, at
\begin{equation} \label{eq:crossEnd}
f = \frac{t+1+\sqrt{(\beta-1)^2 + 2t^2}}{2(\beta+1)},
\end{equation}
giving the crossover a width of
\begin{equation}
\Delta f = \frac{\beta - 2 + 2t + \sqrt{(\beta - 1)^2 + 2t^2}}{2(\beta + 1)}
\end{equation}
as illustrated in Fig~\ref{fig:2x1overlay}.
(Note that this $\Delta f$ is the width of a single transition, in contrast to Eq.~\ref{eq:fluxInjectionWidth} which is the range of frustration over which particles vortices enter the array.)
This should be a good approximation so long as all tunnelling energies remain small compared with interaction energies, $t\ll \beta$.

\begin{figure}
	\includegraphics[width=8cm]{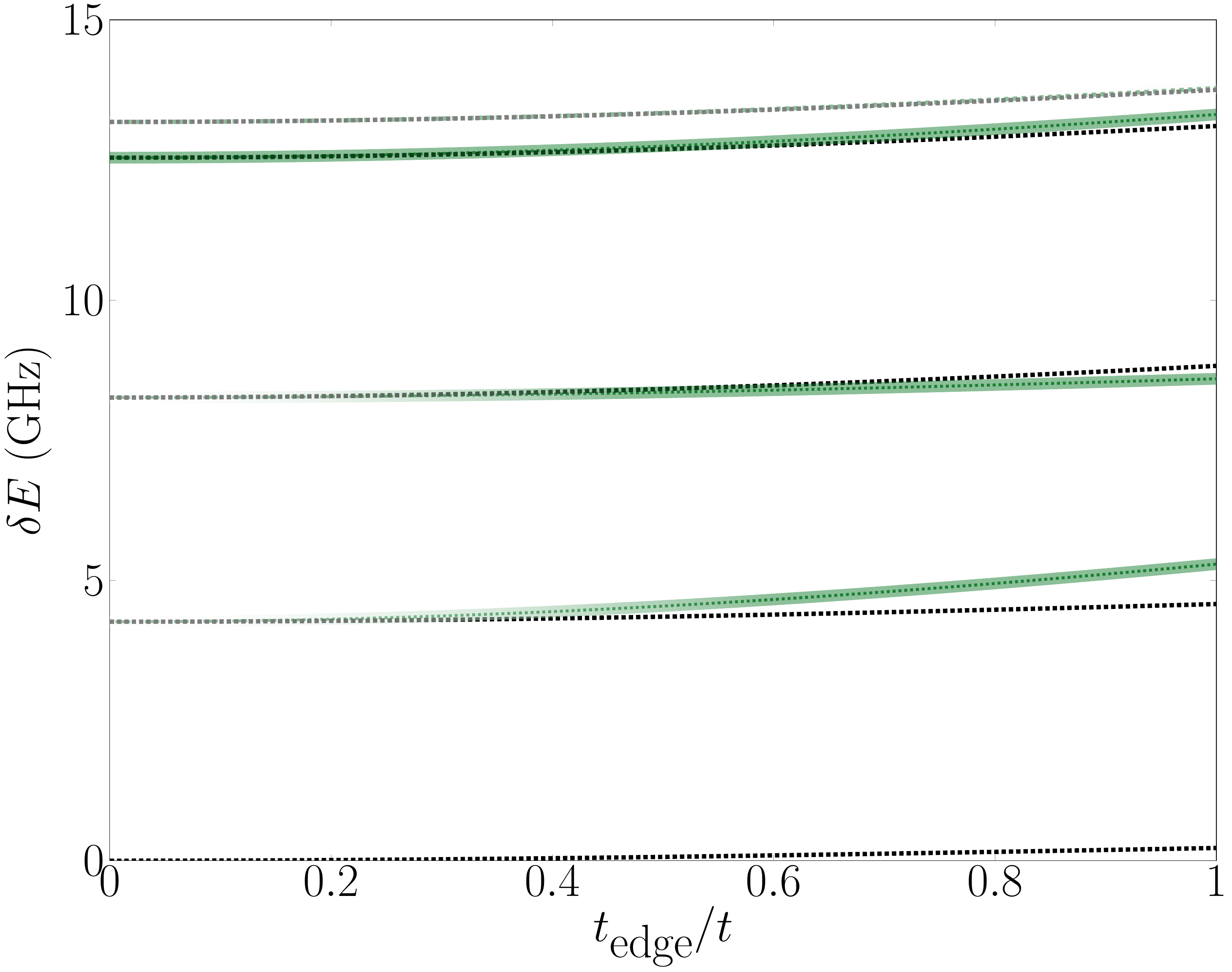}
	\centering
	\caption{ \label{fig:Perturbation} Energy gaps of a two-site system as the junctions on the boundary are turned on with an external frustration of $f = 0.24$, in the centre of the crossover region described by Eqs.~\ref{eq:crossBegin} and \ref{eq:crossEnd}. Thick green lines indicate the matrix element, as in Fig.~\ref{fig:2x1overlay}. Black dotted lines give the results of second-order perturbation theory, given in the Appendix. It can be seen that some levels acquire a finite matrix element as tunnelling through the boundary increases. This occurs as the total number of particles is no longer conserved, and eigenstates of the Hamiltonian consist of superpositions of states of different particle number. }
\end{figure}

We can also calculate the shift in energy levels perturbatively, as the edge-tunnelling is gradually increased from zero, so as to see how the flat bands with sudden transitions for the closed case map smoothly onto the curved bands with gradual crossovers seen in the open case.
Shifts in the energy levels are calculated to second-order, and the resulting gaps are plotted along with the corresponding numerical calculations in Fig.~\ref{fig:Perturbation}.
The change in energy is minimal at most values of the frustration, so we tune the frustration to sit in the middle of the transition where the number of particles in the ground state changes (the precise value chosen is given by the average of Eq.~\ref{eq:crossBegin} and Eq.~\ref{eq:crossEnd}).
A full derivation of these results is given in the appendix.

As we move to larger systems, analytic calculations become impractical even in the closed case.
However, some features from the two-site system will remain generally true.
Systems with no tunnel junctions on the exterior edges will always host states of well-defined particle number, and give rise to response functions which are independent of external frustration except for sharp sudden transitions when the ground-state particle number changes.
Adding exterior tunnel junctions will mean that particle-number is no longer well-defined in general, and will cause all energy gaps and matrix elements - and by extension, the response functions - to be frustration-dependant.
Sharp, sudden transitions will give way to smooth, continuous crossovers.

\section{Including dissipation}
Despite vast advancements in fabrication techniques over the past decades, dissipation is still present any experiment on superconducting networks.
This offers something of a paradox - superconductors do not have any intrinsic resistance, yet resistance is frequently observed in experiment.
The precise origin of dissipation in these systems is contentious.
It persists at temperatures $T \ll \Delta/K_B$, voltages $V \ll \Delta/2e$ and currents far below a junction's critical current, so that quasi-particle effects should be negligible.
Such dissipation is also observed in circuits fabricated from low transparency junctions with negligible sub-gap leakage \cite{Duty}.
Nevertheless, the dissipation in arrays of junctions is there \cite{Duty,Cole2015,Cedergren2015,Chiena,Kafanov2008,Schafer2013}.
To tackle this problem we will need to consider our circuits to be open quantum systems with some dephasing.

There have been many different approaches to generalizing the Kubo formula to open quantum systems \cite{Ban2017,Saeki2010,Uchiyama2009}.
We shall adopt the method presented in \cite{Venuti2016}, which is based upon considering first the Liouvillian of the open quantum system $\mathcal{L}_0$ and then treating the driving force $f(t)$ as a perturbation $f(t)\mathcal{L}_1$.
To simplify things further, we will assume $\mathcal{L}_1$ is of Hamiltonian type (i.e. non-dissipative).
This allows us to write the Kubo formula as
\begin{equation} \label{eq:openKubo}
	\chi_{\phi}(t) = i\theta(t)\textrm{Tr}\left\{[\phi(t),\tilde{\rho}]\phi \right\}
\end{equation}
where $\tilde{\rho}$ is the steady-state density matrix and the time evolution $\phi(t)$ is generated by $\mathcal{L}_0$.
So the calculation of the response for an open system involves first calculating the steady state density matrix $\tilde{\rho}$ defined by $\mathcal{L}_0\tilde{\rho} = 0$, and then calculating the time-evolution of the operator $\phi$ under the action of $\mathcal{L}_0$ (compare to the closed-system case, where we used the ground-state density matrix, and the time evolution of $\phi$ was generated by the Hamiltonian).

Due to the present lack of a complete microscopic model for dissipation in superconducting devices, we treat dissipation phenomenologically.
In the present work we will consider dephasing due to charge and flux noise, however we note other channels of decoherence and loss will also play an important role.

To that end, our Liouvillian $\mathcal{L}_0$ is given by a Lindblad equation \cite{Lindblad1976,Breuer2002}
\begin{equation} \label{eq:Lindblad}
\mathcal{L}_0 = - \frac{i}{\hbar}\left[H_S,\rho\right] + \sum_{k} \Gamma_k\left(L_k\rho L^\dagger_k - \frac{1}{2} \left\{L^\dagger_k L_k, \rho  \right\} \right)
\end{equation}
where $L_{k}$ are the Lindblad operators $Q$ and $\Phi$ for each site in the system, $\Gamma_k$ are dephasing rates and $\rho(t)$ is the density matrix.

We will need to make some assumptions about the coupling of the environment in order to select appropriate Lindblad operators.
Important sources of noise in superconducting circuits are charge and flux fluctuations in the environment \cite{Simmonds2004,Yan2012,Bylander2011,Gustavsson2011}, so it is natural to assign $L_1 = \sum_j \hat{Q}_j/2e$, $L_2 = \sum_j \hat{\Phi}_j/\Phi_0$, where by summing over all sites we are implicitly assuming that the coupling is homogeneous across the device.
Another process to consider would be dissipative quantum tunnelling, which we can include via a Lindblad operator
\begin{equation}
\begin{split} 
L_3 = \sum_{n,m,\langle i,j \rangle}&\left[ \frac{}{} \ket{n_i,m_j}\bra{n_i+1,m_j-1} \right.
\\ 
 &+ \left.  \ket{n_i,m_j}\bra{n_i-1,m_j+1} \right].
\end{split}
\end{equation}
In a circuit with junction boundaries, boundary terms which change the total number of particles in the system may be included in the definition of $L_3$.

\begin{figure}[tb!]
	\centering \includegraphics[width=8cm]{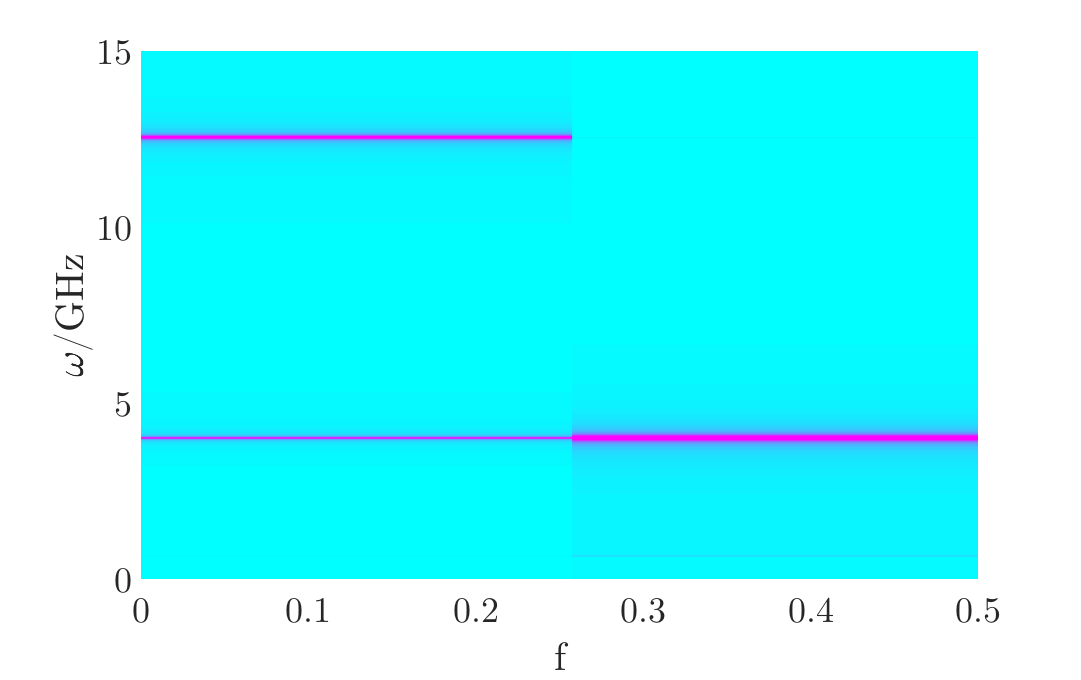}
	
	\vspace{0.8cm}
	
	\centering \includegraphics[width=8cm]{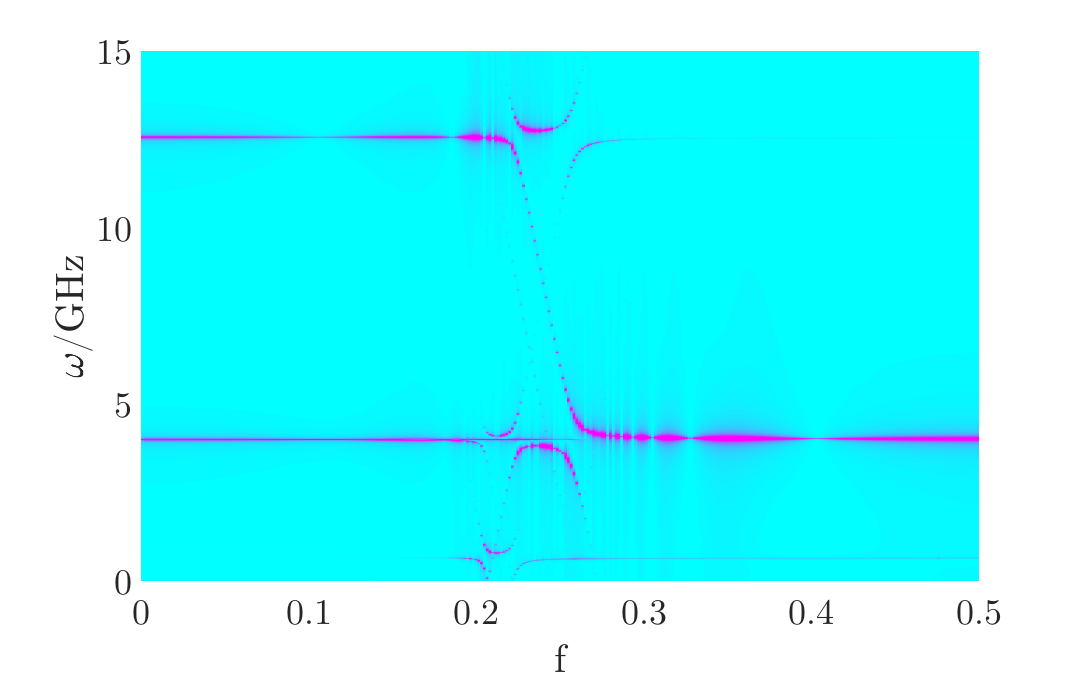}	
	\caption{\label{fig:Dephasing} The dynamical response $\chi_{n}(\omega)$ for a two-site system with closed (top) or open (bottom) edges with dephasing, calculated using Eq.~\ref{eq:openKubo}, with dynamics given by the Lindblad equation Eq.~\ref{eq:Lindblad}. Compared with the (pure) ground state calculations in Fig.~\ref{fig:2SiteClosed} and Fig.~\ref{fig:2SiteOpen} respectively, additional lines appear in the spectrum, corresponding to energy gaps relative to other states appearing in the steady state mixture. Some of these additional states are listed in Fig.~\ref{fig:chargeNoise}.}
\end{figure}

Since we have no microscopic model for the decoherence channels, we select our $\Gamma_k$ phenomenologically.
A reasonable estimate for the minimum dephasing present would come from the inverse dephasing time $1/T_2$ of circuits discussed in the literature.
Transmon qubits in 3D cavities (similar to the cavity systems we consider in the present work) can routinely achieve $T_2 ~ 20 \mu\textrm{s}$ \cite{Paik2011}, which would give us a dephasing rate of the order of $10^{4}$ Hz.
In practice, most many-site devices will fare far worse than the 3D transmon qubit, so we will take $10^4$ Hz as a lower bound and examine the response spectrum as the dephasing rate is increased beyond that.

\begin{figure*}[tb!]
	\centering \includegraphics[width=\linewidth]{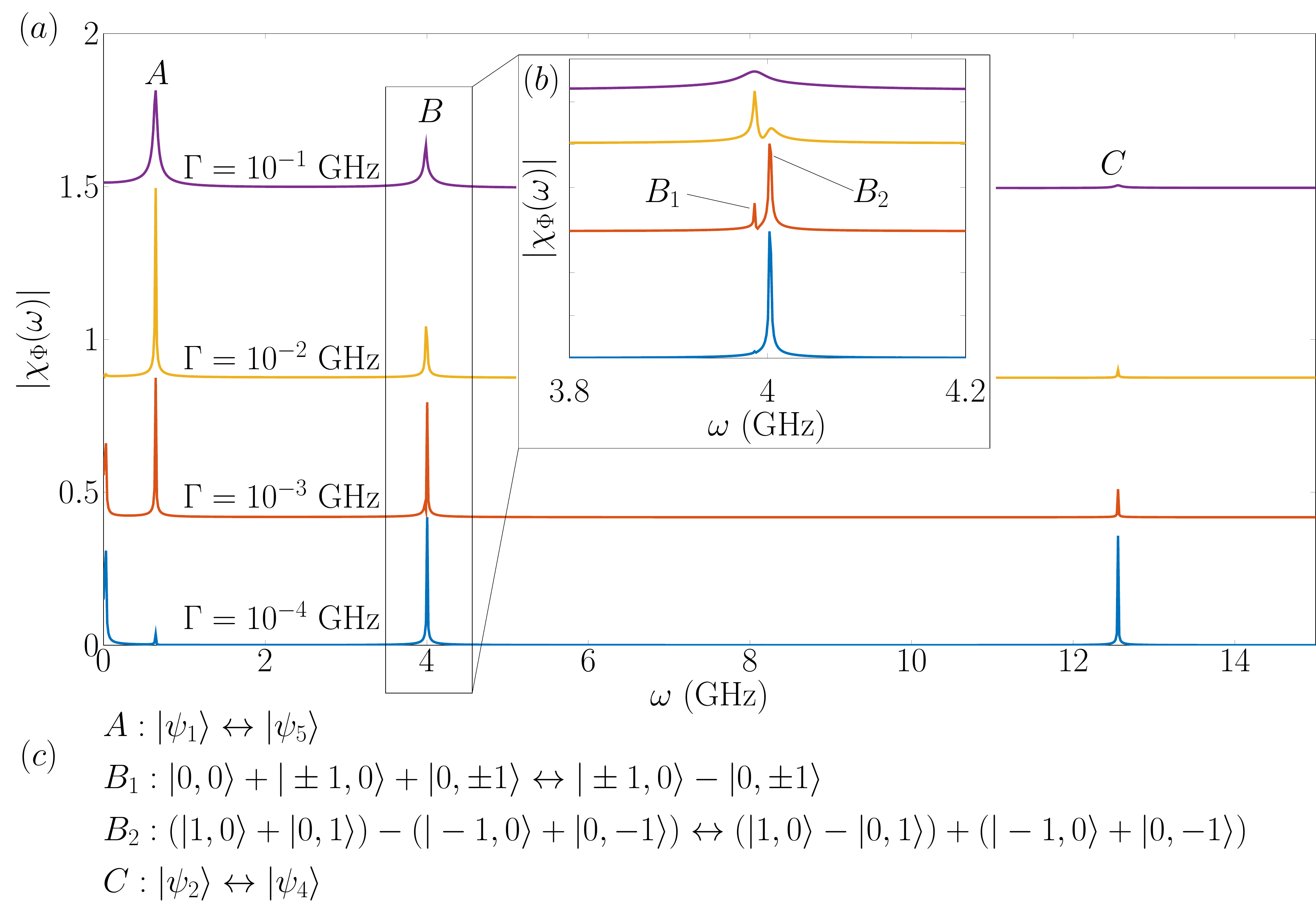}
	\caption{\label{fig:chargeNoise} The linear dynamical response $\chi_n(\omega)$ for $f=0$ as the charge dephasing rate $\Gamma_1$ is adjusted while all other dephasing rates are fixed at 0. The peaks $A$, $B_1$, $B_2$ and $C$ labelled in (a) and (b) correspond to transitions between states listed in (c) (using the notation introduced in Appendix~\ref{appendix:eigen}). At zero dephasing, the only peak present is $C$, corresponding to the transition between the ground and first excited state (see Fig.~\ref{fig:2x1overlay}). Dephasing drives the system from the ground state into a mixed state, so that other transitions can contribute. As the dephasing rate is increased, the peaks broaden until, at strong dephasing, important features are washed out completely. }
\end{figure*}

To solve the Lindblad equation numerically for QPS systems, we will need to represent the charge operator in the basis of flux-number operators.
This is given by 
\begin{equation}
\left(\hat{Q}_j\right)_{nm} = \begin{cases}
\frac{e}{\pi}\left(\frac{1}{2\pi}\right)^N \frac{i(-1)^{\Phi_n - \Phi_m }}{\Phi_n - \Phi_m} \delta_{\tilde{\Phi}_n,\tilde{\Phi}_m}, & n \neq m\\
0, & n=m.
\end{cases}
\end{equation}
where $N$ is the number of sites (here we consider N=2), $\Phi_n$ is the total number of flux quanta in state $n$, and $\tilde{\Phi}_n$ is a vector of the number of flux quanta on every site except $j$ in state $n$.
A derivation of this result is given in Appendix~\ref{ap:chargeOp}.

We present here numerical calculations of the linear dynamical susceptibility $\chi_n(\omega)$ as a function of the external frustration $f$ for a two-site system with both hard- and junction-boundaries presented in Fig.~\ref{fig:Dephasing}, where we have chosen $\Gamma_1 = 10^{-4}$ GHz, $\Gamma_2 = 10^{-2}$ GHz and $\Gamma_3 = 0$, corresponding to charge noise, flux noise and dissipative tunnelling respectively (note that by setting $\Gamma_3 = 0$ we are neglecting dissipative tunnelling and assuming on-site noise to be dominant dephasing pathways).
To illustrate more explicitly the effect of dephasing on the system, we have also calculated the linear response for a fixed frustration $f=0$ as a function of the dephasing rate $\Gamma_1$, while other rates have been fixed to zero, Fig.~\ref{fig:chargeNoise}.
The effect of this dephasing is to drive the system into a mixed state, $\rho_M$.
The response function for the system in this state is given by Eq.~\ref{eq:mixedSpecRep}.
The presence of additional states in the mixture leads to the presence of additional peaks in the response spectrum, while the process of dephasing itself leads to a broadening of the peaks.

In addition to dephasing, a realistic system may also exhibit relaxation.
We have neglected such effects here, as the precise rates depend on both the system eigenvalues and the functional form of the noise spectrum for each noise source.
More sophisticated techniques, such as the Bloch-Redfield master equation \cite{Jeske2013,Lim2017,Redfield1965} may be required for such an undertaking.

\section{$(3\times 2)$-site system}

In a full quantum treatment with exact diagonalization, we are limited to relatively small systems due to the prohibitively large Hilbert space of the problem.
Even if we are able to restrict ourselves to a maximum of $\pm 1$ excitation per site, the size of the Hilbert space scales exponentially with the number of sites.

\begin{figure*}[tb!]
\begin{minipage}[t]{0.4\linewidth}
	\centering{a)}\hbox{\hspace{1.2cm}\includegraphics[width=4.4cm,angle=90,origin=c]{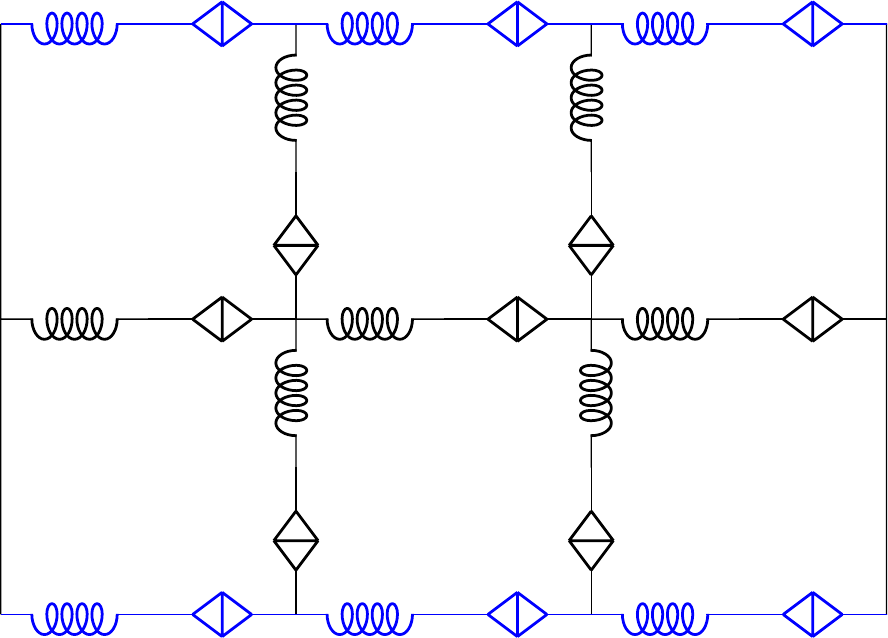}}
\end{minipage}
\hfill
\begin{minipage}[t]{0.48\linewidth}
	\centering{b)}\includegraphics[width=\linewidth]{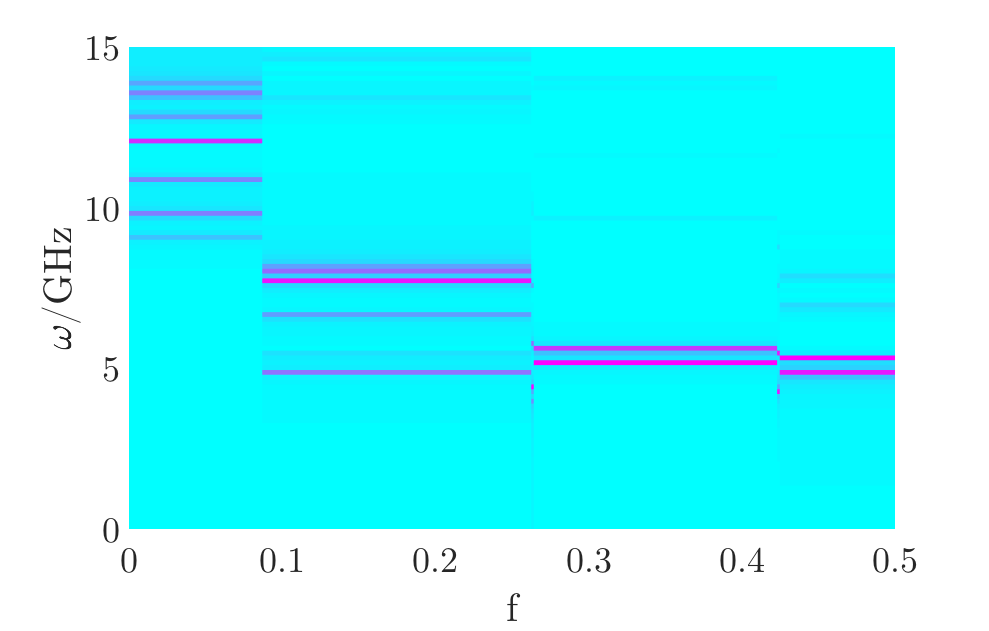}
\end{minipage}
\vfill
\begin{minipage}[t]{0.48\linewidth}
	\centering{c)}\includegraphics[width=0.85\linewidth]{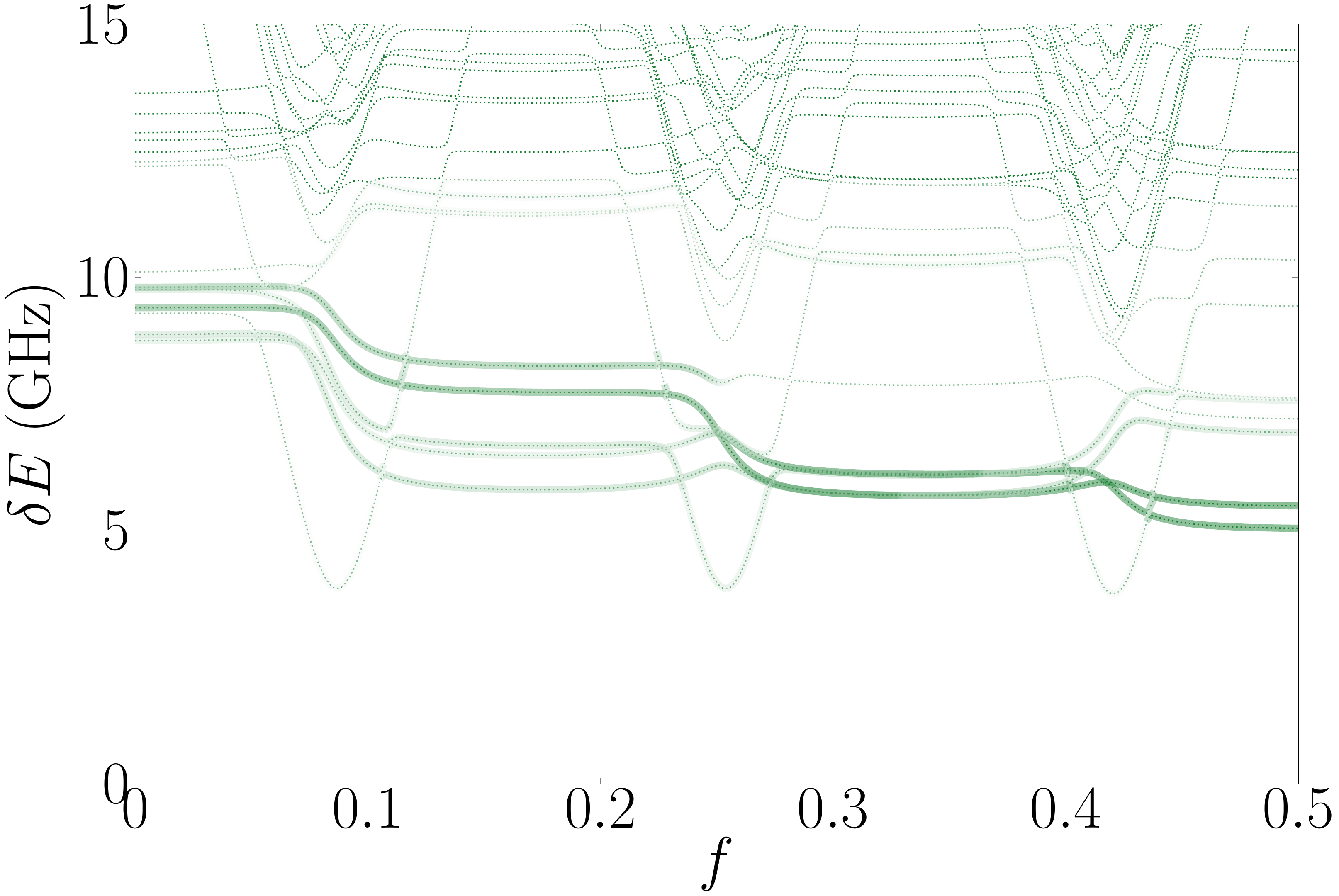}
\end{minipage}
\hfill
\begin{minipage}[t]{0.48\linewidth}
	\centering{d)}\includegraphics[width=\linewidth]{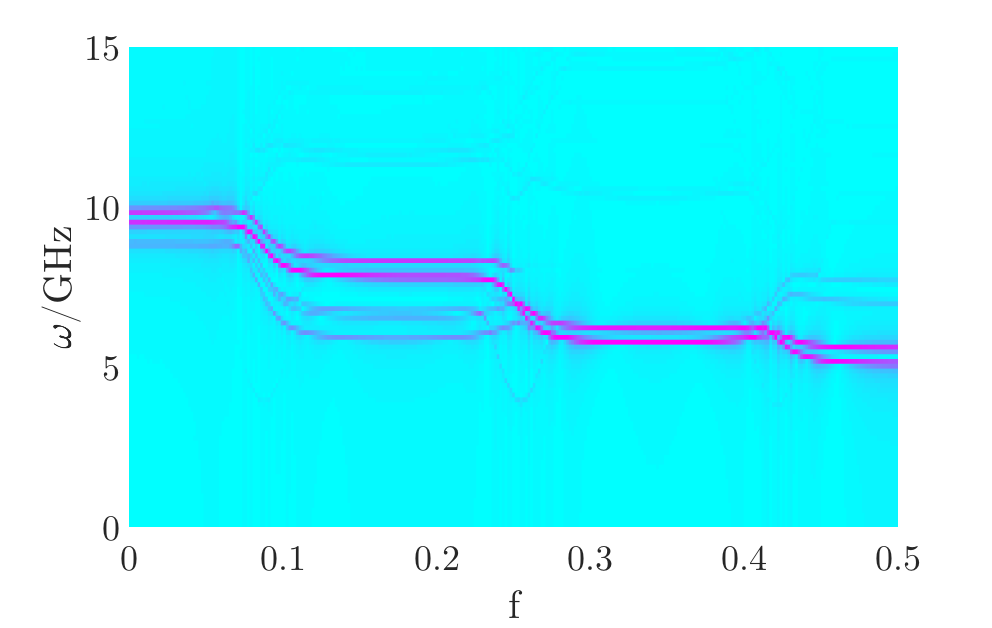}
\end{minipage}
	\caption{\label{fig:6loopQPS} a) Circuit diagrams for a 2$\times$3 loop flux-based circuit. The relevant degrees of freedom for this circuit are vortices in the loops, which may tunnel across the QPS elements on the branches. In the top circuit, total vortex number is conserved, whereas in the bottom circuit vortices may enter and exit the array by tunnelling across the outer edges. b),d) The dynamical response $|\chi_{n}(\omega)|$ for the 2$\times$3 grids shown in Fig.~\ref{fig:6loopQPS} with hard (top) and junction (bottom) boundaries. c) The energy gaps about the ground state energy plotted as a function of external flux (dotted lines). Colour of thick, solid lines corresponds to the amplitude-squared of the matrix element for the vortex number operator between that state and the ground state, $|\bra{\psi_n}\hat{\Phi}_1\ket{\psi_0}|^2$, c.f. Eq.~\ref{eq:specRep}. }
\end{figure*}

We now consider a $2\times 3$ system with circuit diagram depicted in Fig.~\ref{fig:6loopQPS}.
The response spectrum for this circuit is calculated numerically, with the same parameters as the $2\times 1$ calculations.

Despite the increase in complexity and computational cost in larger systems, we see many of the features present in the spectra resemble features present in the more simple 2-site system.

In the spectra for this circuit we see four distinct regions as we vary $f$, corresponding to a total of 0, 1, 2 or 3 particles in the ground state. 
For hard boundary conditions, tuning $f$ causes sharp transitions between regions of different ground-state particle number.
However, when the boundaries contain tunnel junctions the total number of particles in the system is no longer a conserved quantity, and we see smooth, gradual transitions between the different regions.
Within these transition regions, the ground state consists of a superposition of different particle numbers.

The spectra presented in Fig.~\ref{fig:6loopQPS} b) and c) can be understood as arising from Eq.~\ref{eq:specRep}.
The frequency of each of the lines is given by the gap between the ground and excited energy levels, and the height or magnitude of the response is given by the matrix element $|\bra{\psi_i}\hat{n}\ket{\psi_j}|^2$.
In Fig.~\ref{fig:6loopQPS} c), we plot all of the gaps above ground in the junction-boundary system as dashed green lines.
The thick, solid lines appearing in c) also follow the gaps, but with a colour weighted by the matrix element, so that this curve gives the same response spectrum as d).

\begin{figure}
	\includegraphics[width=\linewidth]{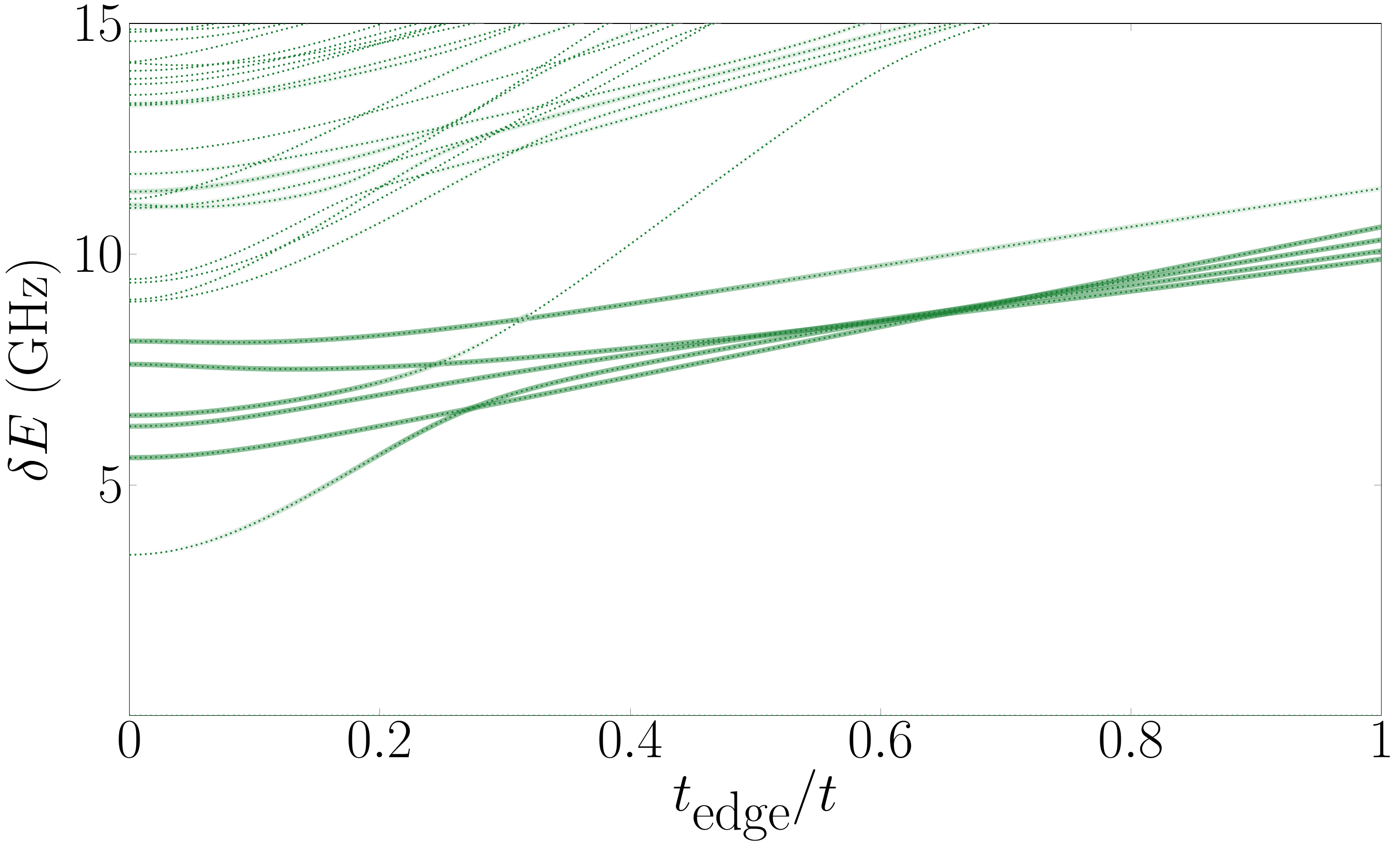}
	\centering
	\caption{ \label{fig:2x3edges} Energy gaps above ground $E_i - E_0$ as a function of the boundary tunnelling $t_\textrm{edge}$. The darkness of the solid green lines indicated the value of the matrix element $|\bra{\psi_i}\hat{n}\ket{\psi_0}|^2$. These lines are not visible where the matrix element vanishes.}
\end{figure}

We can examine the way in which the sharp transitions in the hard-boundary system map onto the smooth transitions in the junction-boundary system by looking at how the energy levels shift and the boundary tunnel amplitude is gradually turned on from zero.
The result is plotted in Fig.~\ref{fig:2x3edges}, where gaps in energy levels $E_i - E_0$ are represented as dashed green lines, and the corresponding matrix elements $|\bra{\psi_i}\hat{n}\ket{\psi_0}|^2$ are represented by the darkness of the thick solid green lines, in a manner analogous to the two-site calculation presented in Fig.~\ref{fig:Perturbation}.

The 2$\times$3 array differs from the 1$\times$2 in that, for each value of $f$, there are many excited states with the same number of particles as the ground state, and therefore many lines in the response spectrum.
This arises simply from having a greater number of different ways to arrange $N$ particles on six sites than on two.
As the number of sites increases, more and more lines will appear in the spectrum, and the gap between them will decrease.
As systems approach a large number of sites, these distinct spectral lines will merge together in a manner analogous to the formation of energy bands in solids.
Indeed, in experiments on 2D Josephson junction arrays consisting of 90 loops, the measured response spectrum forms a single band \cite{Cosmic2018a}.

\section{Conclusion}

Spectroscopy experiments in microwave cavities provide a new and fruitful avenue for studying the dynamics of superconducting networks while minimising the additional noise due to the measurement apparatus itself.
Here we have explored linear response theory as a theoretical tool to connect circuit theoretic formulations of superconducting networks to microwave spectroscopy experiments.

Characteristic features of these spectra as the external frustration $f$ is varied correspond to changes in the number of particles in the ground state.
This can be compared with a similar situation with much larger Josephson junction array system, where features in the experimentally obtained response spectrum corresponded to changes in total number of vortices in the array \cite{Cosmic2018a}.

Decoherence and dissipation were given only a cursory treatment here, and further work will investigate this in more detail.
In particular, the present work focused only on dephasing via a Lindblad formalism, and only for the simple case of a two-site system.
An obvious next step would be a more thorough and sophisticated treatment of decoherence, in particular relaxation processes.

We were able to understand the key features of the response spectrum in terms of the spectral representation Eq.~\ref{eq:specRep}.
Analytic results were derived for $2\times 1$ circuits, and many of the features exhibited by these simple cases have counterparts in the larger $2\times 3$ circuits which we studied numerically.

We focused here on small systems which were amenable to solution via exact diagonalisation.
This gives us insight into the effects of the boundary of the system, and allows us to identify signatures of changes in the number of vortices in the ground state of the system.
These results will also be important for benchmarking the approximation schemes which will be necessary for treating larger systems.

\section{Acknowledgements}
We thank N. Vogt, F. Hassler, R. Cosmic, Y. Nakamura, H. Ikegami and C. M\"{u}ller for useful discussions. This work was supported in part by the Australian Research Council under the Discovery and Centre of Excellence funding schemes (project numbers DP140100375 and CE170100039). Computational resources were provided by the NCI National Facility systems at the Australian National University through the National Computational Merit Allocation Scheme supported by the Australian Government.

\bibliography{Bib}

\appendix
\section{Eigenstates and eigenvalues of Eq.~\ref{eq:twoSite}} \label{appendix:eigen}
The Hamiltonian in Eq.~\ref{eq:twoSite} can be exactly diagonalized, and we find that te eigenstates are
\begin{equation} \label{eq:exactStates}
\begin{split}
\ket{\psi_1} =& \frac{1}{\sqrt{2}}\left( \ket{-1,1} - \frac{}{} \ket{1,-1}\right)\\
\ket{\psi_2} =& \ket{-1,-1}\\
\ket{\psi_3} =& \ket{1,1}\\
\ket{\psi_{4,5}} =& \mathcal{N}_{4,5} \left[\ket{-1,1} + \ket{1,-1} \frac{}{} \right. \\ 
&+ \left. \frac{\beta - 1 \pm \sqrt{ (1-\beta)^2 + 2 t^2}}{t}\ket{0,0} \right]\\
\ket{\psi_{6,7}} =& \frac{1}{\sqrt{2}}\left( \ket{0,-1} \pm \frac{}{} \ket{-1,0} \right)\\
\ket{\psi_{8,9}} =& \frac{1}{\sqrt{2}} \left( \ket{0,1} \pm \frac{}{} \ket{1,0} \right)
\end{split}
\end{equation}       
with eigenvalues
\begin{equation} \label{eq:exactEnergies}
\begin{split}
\lambda_1 =& (\beta -1) + (\beta + 1)f^2\\
\lambda_{2,3} =& (\beta + 1)(f \mp 1)^2\\
\lambda_{4,5} =& \frac{1}{2}(\beta - 1) + (\beta + 1)f^2 \mp \frac{1}{2}\sqrt{(\beta-1)^2 + 2t^2}\\
\lambda_{6,7} =& f(f-1)(\beta+1) + \frac{1}{2}( \beta \pm t )\\
\lambda_{8,9} =& f(f+1)(\beta+1) + \frac{1}{2}( \beta \pm t )\\
\end{split}
\end{equation}
where $\mathcal{N}_{4,5}$ are normalization constants.

\section{Flux-basis representation of charge operator} \label{ap:chargeOp}
For numerical calculations with our vortex-lattice quantum phase model, it is necessary to represent charge $Q$ in the vortex-number basis. To do this we note that within the quasicharge approximation, eigenvalues of $Q$ are restricted to the interval $(-e,+e)$ (otherwise energy can be lowered by tunnelling of a single Cooper-pair - remember, the quasicharge approximation requires that the microscopic degrees of freedom are in their ground states at any point in time). 
Thus charge acts like a phase variable, while flux - which we are so used to thinking of a phase - acts like a particle number (in units of $e=\hbar=1$, we have $\hat{\Phi} = \pi\hat{N}$ where $\hat{N}$ is the vortex-number operator).
To express the charge in terms of a more familiar phase operator (with period $2\pi$ instead of $2e$) we have $\hat{Q} = (e/\pi)\hat{Q}'$.
The relationship between charge and flux in this approximation is just like other phase-number relationships found throughout quantum mechanics.
In particular, we have
\begin{equation}
\braket{\Phi}{Q} = \frac{1}{\sqrt{2\pi}}e^{-i\Phi Q/\hbar}.
\end{equation}
When can therefore express the matrix elements of the charge operator in the vortex-number basis as
\begin{widetext}
	\begin{equation}
	\begin{split}
	\left(\frac{\pi}{e}\hat{Q}_j\right)_{nm} =& \bra{\vec{\Phi}_n}\hat{Q}'_j\ket{\vec{\Phi}_m} = \int_{-\pi}^{+\pi} \textrm{d}\vec{Q}' \braket{\vec{\Phi}_n}{\vec{Q}'}\bra{\vec{Q}'}\hat{Q}'_j\ket{\vec{\Phi}_m}
	= \left(\frac{1}{2\pi}\right)^N \int_{-\pi}^{+\pi} \textrm{d}\vec{Q}' Q'_j e^{i(\vec{\Phi}_m - \vec{\Phi}_n)\cdot\vec{Q}}\\
	=& \left(\frac{1}{2\pi}\right)^N\left[ \int_{-\pi}^{+\pi}\textrm{d}Q'_j Q'_j e^{i(\Phi_j^m - \Phi_j^n)Q'_j} \right]\left[\int_{-\pi}^{+\pi}\textrm{d}\tilde{Q}' e^{i(\tilde{\Phi}_m - \tilde{\Phi}_n)\cdot \tilde{Q}'} \right]\\
	=& \left(\frac{1}{2\pi}\right)^N \left[ \frac{2i\sin[\pi(\Phi_j^m - \Phi_j^n)] - \pi(\Phi_j^m - \Phi_j^n)\cos[\pi(\Phi_j^m - \Phi_j^n)]}{(\Phi_j^m - \Phi_j^n)^2}  \right]\delta_{\tilde{\Phi}_n,\tilde{\Phi}_m}
	\end{split}
	\end{equation}
\end{widetext}
here we have introduced the notation that $\vec{Q}$ ($\vec{\Phi}$) is the vector of the charge (flux) operator for each site in the lattice, and $\tilde{Q}$ ($\tilde{\Phi}$) is the vector of charge (flux) operators for every site \textit{except} $j$.
In integrating over $\tilde{Q}$ we have used the fact that $\Phi_m - \Phi_n$ is always integer, and thus the integral is zero unless $\tilde{\Phi}_m = \tilde{\Phi}_n$ (note that this does not necessarily imply that $\Phi_j^n = \Phi_j^m$).
Defining $\varphi_{nm} \equiv \Phi_j^m - \Phi_j^n$, we can use the fact that vortex numbers are always integer to simplify this further:
\begin{equation}
\left(\frac{\pi}{e}\hat{Q}_j\right)_{nm} = \left(\frac{1}{2\pi}\right)^N \frac{i(-1)^{\varphi_{nm}}}{\varphi_{nm}} \delta_{\tilde{\Phi}_n,\tilde{\Phi}_m}
\end{equation}
except when $\varphi_{nm} = 0$, in which case the integral over $Q'_j$ is zero, so all diagonal elements of $Q'_{nm}$ are zero.

\section{Second-order perturbtion theory.} \label{ap:pert}
We wish to find the leading-order corrections to the eigenvalues in Eq.~\ref{eq:exactEnergies} due to perturbations in the form of Eq.~\ref{eq:boundary},
\begin{equation}
	\lambda_n = \lambda_n^{(0)} + \lambda_n^{(1)} + \lambda_n^{(2)} + \mathcal{O}(t_\textrm{edge}^3),
\end{equation}
where $\lambda_n^{(0)}$ are the exact hard-boundary eigenvalues given by Eq.~\ref{eq:exactEnergies}.
We shall proceed using the standard techniques of time-independent perturbation theory (see, for example, \cite{Sakurai2014}).

The first order term vanishes, because the matrix element $\bra{\psi_n}\hat{V}\ket{\psi_m}$ is zero when $\ket{\psi_n}$ and $\ket{\psi_m}$ are superpositions of states with a fixed number of particles $N$.
However, the eigenstates \textit{do} shift at first order, attaining contributions from states of different numbers of particles.
This means that the matrix element $\bra{\psi_n^{(0)}}\hat{V}\ket{\psi_m^{(1)}}$ may be non-zero, and the energy levels will shift at second order.
When levels are non-degenerate, we can calculate the change in energy via the standard formula from second-order perturbation theory \cite{Sakurai2014}
\begin{equation}
\lambda_n^{(2)} = \sum_{m\neq n} \frac{\bra{\psi_m^{(0)}} \hat{V} \ket{\psi_n^{(0)}} }{\lambda_n^{(0)} - \lambda_m^{(0)}}.
\end{equation}
Using the eigenstates and eigenvalues given by Eq.~\ref{eq:exactStates} and Eq.~\ref{eq:exactEnergies}, we obtain
\begin{equation}
\begin{split}
\lambda_1^{(2)} =& t_\textrm{edge}^2 \left[\frac{1}{\lambda_1^{0} - \lambda_6^{0}} + \frac{1}{\lambda_1^{0} - \lambda_8^{0}}\right]\\
\lambda_2^{(2)} =& \frac{2t_\textrm{edge}^2}{\lambda_2^{0} - \lambda_6^{0}}\\
\lambda_3^{(2)} =& \frac{2t_\textrm{edge}^2}{\lambda_3^{(0)} - \lambda_8^{(0)}}\\
\lambda_4^{(2)} =& 2t_\textrm{edge}^2 \left(\frac{1 + \mathcal{A}_4}{\sqrt{2 + \mathcal{A}_4^2}}\right)^2\left[\frac{1}{\lambda_4^{(0)} - \lambda_6^{(0)} } + \frac{1}{\lambda_4^{(0)} - \lambda_8^{(0)} }\right]\\
\lambda_5^{(2)} =&  2t_\textrm{edge}^2 \left(\frac{1 + \mathcal{A}_5}{\sqrt{2 + \mathcal{A}_5^2}}\right)^2\left[\frac{1}{\lambda_5^{(0)} - \lambda_6^{(0)} } + \frac{1}{\lambda_5^{(0)} - \lambda_8^{(0)} }\right]\\
\lambda_6^{(2)} =& t_\textrm{edge}^2\left[\frac{1}{\lambda_6^{(0)} - \lambda_2^{(0)} } + \left(\frac{1+\mathcal{A}_4}{\sqrt{2 + \mathcal{A}_4^2 }}\right)^2 \frac{1}{\lambda_6^{(0)} - \lambda_4^{(0)}} \right. \\
& \left. + \left(\frac{1+\mathcal{A}_5}{\sqrt{2 + \mathcal{A}_5^2}}\right)^2 \frac{1}{\lambda_6^{(0)} - \lambda_5^{(0)}}   \right]\\
\lambda_7^{(2)} =& \frac{2t^2}{\lambda_7^{(0)} - \lambda_1^{(0)}}\\
\lambda_8^{(2)} =& t_\textrm{edge}^2\left[\frac{1}{\lambda_8^{(0)} - \lambda_3^{(0)}} + \left(\frac{1 + \mathcal{A}_4}{\sqrt{2 + \mathcal{A}_4^2}}\right)^2\frac{1}{\lambda_8^{(0)} - \lambda_4^{(0)} }\right. \\
& \left. + \left(\frac{1 + \mathcal{A}_5}{\sqrt{2 + \mathcal{A}_5^2}}\right)^2\frac{1}{\lambda_8^{(0)} - \lambda_5^{(0)} }  \right]\\
\lambda_9^{(2)} =& \frac{2t^2}{\lambda_9^{(0)} - \lambda_1^{(0)}}
\end{split}
\end{equation}
where
\begin{equation}
	\mathcal{A}_{4,5} = \frac{\beta - 1 \pm \sqrt{ (1-\beta)^2 + 2 t^2}}{t}.
\end{equation}

\section{Dual circuits} \label{app:Dual}
In this paper we have discussed JJAs in a limit where each JJ can be approximated by a QPS element. This is known as a passive duality -- they physical circuit is not changed, but one element is approximated by its electromagnetic dual. There also exists an active duality -- a different physical circuit which obeys the same dynamical laws. This kind of duality transformation is common practice in electrical engineering \cite{Cherry1949,Pointon1991}, and proceeds according to a set of well-established rules. For a planar circuit described by a graph $\mathcal{G}$, the dual circuit is simply described by the dual graph $\mathcal{G}*$ \cite{Harary1994}, and the circuit elements transform according to $L \longleftrightarrow C$, $V \longleftrightarrow I$.

\begin{figure}
\centering{a)} \includegraphics[width=5cm]{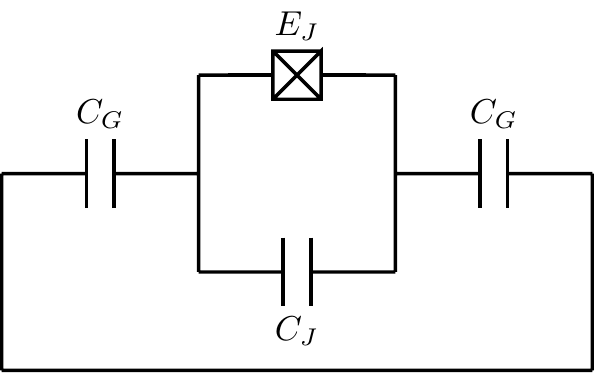}

\vspace{0.8cm}

\centering{b)} \includegraphics[width=7cm]{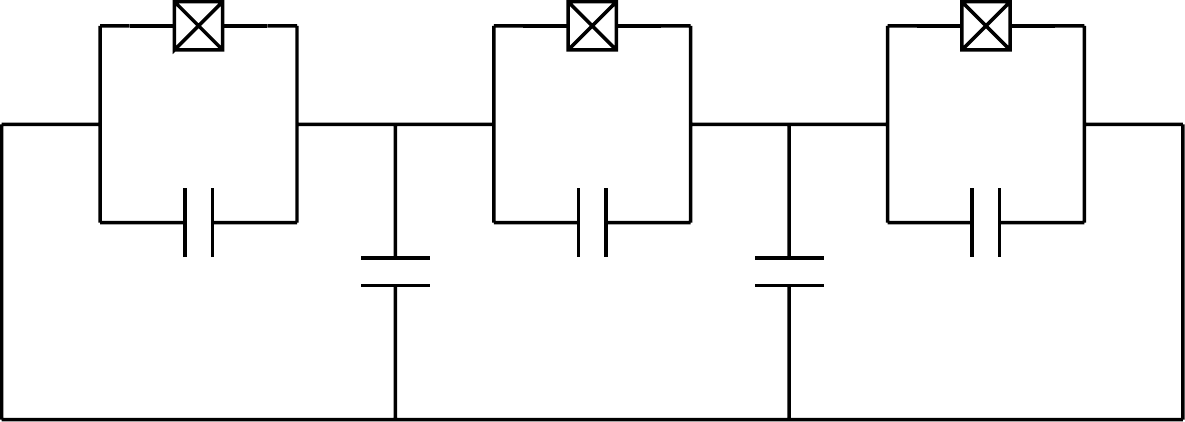}

\vspace{0.8cm}

\caption{\label{fig:dualCircuits} Active duals of the circuits in Figs. 3 and 4 respectively, obtained via a duality transformation. These circuits exhibit the same dynamics as	their duals, but with different variables.  }
\end{figure}

Fig.~\ref{fig:dualCircuits} show the active duals of the circuits in Figs.~\ref{fig:2SiteClosed}, \ref{fig:2SiteOpen} and \ref{fig:2x3edges}. Kinetic inductances $L_K$ are replaced with junction capacitances $C_J$, geometric inductances $L_G$ are replaced with ground capacitances $C_G$, and the effective QPS elements $E_S$ are replaced with JJs. The response spectra for these circuits are the same as those given The response spectra for these
circuits are the same as those given Figs. 3, 4 and 9, except that these are now charge susceptibilities $\chi_Q$, rather
than flux susceptibilities $\chi_\Phi$, and the circuit parameters are now $E_J = 1$ GHz, $C_G = 10^{-3}$ nF and $C_J = 10^{-2}$ nF.

\end{document}